\documentclass[journal]{IEEEtran}

\usepackage{xcolor,soul,framed} %,caption

\colorlet{shadecolor}{yellow}
\usepackage[pdftex]{graphicx}
\graphicspath{{../pdf/}{../jpeg/}}
\DeclareGraphicsExtensions{.pdf,.jpeg,.png}

\usepackage[cmex10]{amsmath}
\usepackage{array}
\usepackage{mdwmath}
\usepackage{mdwtab}
\usepackage{eqparbox}
\usepackage{url}
\usepackage{multirow} 
\usepackage{subcaption}
\usepackage{orcidlink}
\usepackage[normalem]{ulem}

\newcommand{\mat}[1]{{\overline{\textbf{#1}}}}
\newcommand{\vg}[1]{{\boldsymbol{#1}}}
\renewcommand{\v}[1]{{\textbf{#1}}}

\begin{document}
\bstctlcite{IEEEexample:BSTcontrol}
    \title{SesQ: A Surface Electrostatic Simulator for Precise Energy Participation Ratio Simulation in Superconducting Qubits}

  \author{
  Ziang Wang\orcidlink{0009-0002-4931-7781}, 
  Shuyuan Guan\orcidlink{0009-0007-1280-7425},
  Feng Wu\orcidlink{0000-0003-1652-9243}, 
  Xiaohang Zhang\orcidlink{0000-0003-2664-0877},
  Qiong Li\orcidlink{0000-0002-8627-4066},
  Jianxin Chen\orcidlink{0000-0002-9365-776X},
  Xin Wan, 
  Tian Xia\orcidlink{0000-0001-9136-6638}, 
  Hui-Hai Zhao\orcidlink{0000-0001-7075-8325}
\thanks{Corresponding authors: Jianxin Chen (e-mail chenjianxin@tsinghua.edu.cn), Tian Xia (e-mail tianxia.ui@gmail.com) and Hui-Hai Zhao (e-mail zhaohuihai@iqubit.org)}
\thanks{F. Wu, S. Guan and H.-H. Zhao are with Zhongguancun Laboratory, Beijing, China.}
\thanks{Z. Wang and X. Wan are with Zhejiang Institute of Modern Physics and Zhejiang Key Laboratory of Micro-nano Quantum Chips and Quantum Control, Zhejiang University, Hangzhou 310027, China.}
\thanks{S. Guan and Q. Li are with School of Cyberspace Science, Faculty of Computing, Harbin Institute of Technology, Harbin 150000, China.}
\thanks{J. Chen is with Department of Computer Science and Technology, Tsinghua University, Beijing 100084, China.}
\thanks{T. Xia is an independent researcher. }
\thanks{X. Zhang is with the Institute for Brain Research, Advanced Interfaces and Neurotechnologies, Shenzhen Medical Academy of Research and Translation, Shenzhen, 518100 China.}
}

% ====================================================================
\maketitle

% === ABSTRACT ====================================================================
% =================================================================================
\begin{abstract}
An accurate and efficient numerical electromagnetic model for superconducting qubits is essential for characterizing and minimizing design-dependent dielectric losses. The energy participation ratio (EPR) is a commonly adopted metric for evaluating these losses, 
but its calculation presents a severe multiscale computational challenge. The conventional finite element method (FEM) requires 3D volumetric meshing, leading to prohibitive computational costs and memory requirements when attempting to capture singular electric fields at nanometer-thin material interfaces. To address this bottleneck, we propose SesQ, a surface integral equation simulator tailored for the precise simulation of the EPR.
By discretizing 2D surfaces, deriving a semi-analytical multilayer Green's function, and employing a dedicated non-conformal boundary mesh refinement scheme, SesQ accurately resolves singular edge fields without an explosive increase in the number of unknowns.
Validations with analytically solvable models demonstrate that SesQ accelerates capacitance extraction by roughly two orders of magnitude compared to commercial FEM tools. While achieving comparable accuracy for capacitance extraction, SesQ delivers superior precision for EPR calculations.
Simulations of practical transmon qubits further reveal that FEM approaches tend to significantly underestimate the EPR. Finally, the high efficiency of SesQ enables rapid iteration in layout optimization, as demonstrated by minimizing the EPR of the qubit pattern, establishing SesQ as a powerful tool for the automated design of low-loss superconducting quantum circuits.
\end{abstract}

% === KEYWORDS ====================================================================
% =================================================================================

% \fi
\IEEEpeerreviewmaketitle

% ====================================================================

% === I. INTRODUCTION =============================================================
% =================================================================================
\section{Introduction}

The performance of a superconducting quantum processor is fundamentally governed by two critical metrics: the number of qubits and the fidelity of quantum operations. 
Achieving high fidelities, which is crucial for fault-tolerant quantum computing, relies heavily on high-coherence qubits. Therefore, extending the coherence time by carefully minimizing energy losses is a central task for improving quantum hardware performance. 

Among the various implementations of superconducting qubits, the transmon has emerged as the most widely adopted qubit architecture~\cite{Koch2007}. The dominant decoherence source of a transmon qubit is the surface dielectric loss, which originates from the two-level system (TLS)~\cite{Martinis2005,Wang2015,Muller2019} defects residing within the nanometer-thin contaminated substrate-metal (SM), metal-air (MA), and substrate-air (SA) interfaces~\cite{Gambetta2017}. Therefore, the ability to accurately predict and minimize the coupling between the qubit's electric field and these lossy interfacial regions is a critical task in the quantum chip design.

The macroscopic quantity used to model this coupling is the energy participation ratio (EPR)~\cite{Wang2015,Minev2021}, which is defined as the fraction of the electric energy within the interfaces to the total energy of the electric field in the whole space. The TLS defects are known to absorb qubit energy via the interactions between the qubit and the oscillating electric field~\cite{Lisenfeld2019}. 
The primary approach to enhance the interface quality is through improving the fabrication processes to reduce the density of TLSs. 
However, under certain fabrication conditions, one can also adjust the footprint or shape~\cite{Wang2015,Eun2023} of the superconductors to minimize the TLS excitation by reducing the EPR.
Extensive experimental studies have verified that changes in the EPR—often achieved by systematically varying the geometries of superconducting qubits and resonators—can significantly affect the coherence time of the device~\cite{Wang2015,Gambetta2017,Gao2008,Dial2016,Ganjam2024}.
Consequently, an accurate simulation of the EPR is essential for low-loss chip design. However, this simulation presents a severe multiscale computational challenge. It requires calculating the electric field energy within interface layers of thickness less than a few nanometers, while qubit dimensions span hundreds of micrometers. Compounding this difficulty, the electric field exhibits a singularity at the edges of superconducting films, where the interfacial energy density reaches its maximum~\cite{Wenner2011,Martinis2022}. Hence, it is essential to bring an accurate multiscale electromagnetic simulator into the design and simulation of qubits~\cite{Elkin2025,Wu2025}. 

Previous approaches to estimating the EPR generally fall into two categories, each with distinct limitations. The first category involves analytical modeling, such as the conformal mapping techniques demonstrated in~\cite{Murray2018,Murray2020}. While these methods provide rapid calculations and physical intuition for standard structures like the coplanar waveguide (CPW), they are difficult to generalize to the complex and arbitrary geometries of typical qubit designs. 
The second category relies on numerical simulators, primarily the Finite Element Method (FEM). While FEM can handle arbitrary geometries, it may fail to accurately simulate the nanometer-scale interfaces. Capturing the singular fields at the edges requires an exponentially dense volumetric mesh, which leads to prohibitive computational costs and memory requirements. 
Ref.~\cite{Wang2015} proposed a manual divide-and-conquer strategy with heuristic scaling to simulate the EPR for various shapes of transmon qubits, which has been adopted by a few subsequent numerical studies~\cite{Eun2023,Park2023,Ganjam2024,Smirnov2024}. 
This procedure involves dividing the original problem into a few $2$D and $3$D FEM simulations with special treatments for boundary singularities. These separate solutions are then combined by matching the boundary conditions to form an overall solution. 
In this manner, the multiscale problem is divided into a few mono-scaled problems, which helps to substantially reduce the computational costs and memory requirements. 
However, the division scheme can be arbitrary or design-dependent, and the combination of the separate solutions is also prone to uncontrollable errors. 
Alternatively, if the FEM is directly applied to arbitrary qubits for $3$D simulations, we found that
it is difficult to suppress the errors to below $10\%$ with affordable simulation costs. 

In this paper, we present a solution to this simulation bottleneck: SesQ, a surface integral equation simulator tailored for the precise simulation of the EPR. 
Unlike the FEM, which requires meshing the entire $3$D volume, our approach utilizes the Surface Integral Equation (SIE) method. This reduces the problem dimensionality, requiring discretization only on the quasi-$2$D superconducting layers. By combining this with a dedicated non-conformal boundary mesh refinement scheme, we can capture the singular electric field at conductor edges with high precision without the significant growth in the number of unknowns. As a result, the EPR simulation problems, usually of a multiscale nature, can be addressed by SesQ in a single-step simulation, which avoids manual intervention and uncontrollable errors in the divide-and-conquer scheme.
We demonstrate that SesQ offers a dual advantage of speed and accuracy. 
It accelerates both capacitance extraction and the calculation of the EPR by orders of magnitude compared to the FEM. While achieving comparable accuracy for capacitance extraction, SesQ delivers superior precision for the EPR calculations, enabling the efficient optimization of superconducting chip layout designs to minimize the dielectric loss.

The remainder of this paper is organized as follows. Section~\ref{sec:method} details the methods, including the formulation of the integral equation method, the derivation of the Green's function for multilayer space, and the mesh refinement techniques. Section~\ref{sec:results} presents numerical results, including validation against analytical solutions, comparisons with FEM for capacitance extraction and EPR calculation, EPR evaluation of realistic qubit designs, and qubit pattern optimization for EPR minimization. Finally, Sec.~\ref{sec:conclusion} concludes this paper.

\section{Method} \label{sec:method}
\subsection{Integral Equation Method}
The integral equation method, also known as the method of moments (MoM), is widely used in computational electromagnetics~\cite{Harrington1993,Gibson2014}. 
The MoM transforms the governing integral equations into a system of linear equations by discretizing the unknown physical quantities—such as surface charge density in electrostatics—into a finite set of basis functions. A significant advantage of the MoM is that it requires discretization only on the surfaces where the sources reside, rather than throughout the entire 3D volumetric domain. This effectively reduces the dimensionality of the problem and significantly decreases the number of unknowns.

This method is particularly useful for electromagnetic problems in open environments because the Green's function can be found analytically~\cite{Chew1995}. 

\subsection{Governing Equation at the Electrostatic Limit}

At the electrostatic limit ($\omega \to 0$), 
the equation that governs the relation between the charge density distribution 
$\varrho (\v{r})$
and the scalar potential $\phi(\v{r})$ is given by
\begin{equation}\label{scalar_helmholtz_eq}
  \nabla \cdot \left[ \epsilon_r(\v{r}) \nabla \phi(\v{r}) \right]
  = -\frac{1}{\epsilon_0} \varrho(\v{r}),
\end{equation}
where $\epsilon_0$ is the permittivity of free space, 
and $\epsilon_r(\v{r})$ is the inhomogeneous relative permittivity of the medium. 

The Green's function for the governing equation~\eqref{scalar_helmholtz_eq}, 
denoted by $G(\v{r}, \v{r}^\prime)$, is the potential response to a source point $\v{r}^\prime$ and is the solution to the following equation:
\begin{equation}\label{scalar_helmholtz_delta_func}
  \nabla \cdot \left[ \epsilon_r(\v{r}) \nabla G(\v{r}, \v{r}^\prime) \right] 
  = -\frac{1}{\epsilon_0} \delta(\v{r}- \v{r}^\prime),
\end{equation}
where $\delta(\v{r}- \v{r}^\prime)$ is the Dirac delta function. 

In free space, the Green's function has a simple analytical form: 
\begin{equation}\label{free_space_green_func}
G(\v{r}, \v{r}^\prime) = \frac{1}{4 \pi \epsilon_0|\v{r}-\v{r}^\prime|}.
\end{equation}
In a two-layer structure, with $\epsilon_r = \epsilon_1$ for $z>0$ and $\epsilon_r = \epsilon_2$ for $z<0$, and 
enforcing that the source points $\v{r}^\prime$ reside only on the $z=0$ plane, the Green's function can be derived using the method of image charges, which yields:
\begin{equation}\label{two_layer_green_func}
G(\v{r}, \v{r}^\prime) = \frac{1}{4 \pi \epsilon_0 \epsilon^{\text{eff}} |\v{r}-\v{r}^\prime|},
\end{equation}
where the effective relative permittivity is $\epsilon^{\text{eff}} = {(\epsilon_1 + \epsilon_2)}/{2}$. 
The above Equations~\eqref{free_space_green_func} and~\eqref{two_layer_green_func} are two of 
the simplest forms of the Green's function for such open environments. 
Alternatively, we will derive other useful forms of the Green's function in Section~\ref{sec:multilayer_green_function}.
% Alternatively, one can find other useful forms of the Green's function analytically or numerically.

Applying Green's second identity to \eqref{scalar_helmholtz_eq} and \eqref{scalar_helmholtz_delta_func} and restricting the solution domain to the surface, we obtain the governing surface integral equation as:
\begin{equation}\label{integral_form}
  \int_\mathcal{S} d^2\v{r}^\prime G(\v{r}, \v{r}^\prime) q(\v{r}^\prime) = \phi(\v{r}),
\end{equation}
where the integration is over the superconductor surfaces $\mathcal{S}$,
and $q (\v{r}^\prime)$ is the surface charge density at the source point $\v{r}^\prime$.
Equation~\eqref{integral_form}
simply indicates that the scalar potential is the summed contribution from all the charges on the superconductors.
In a realistic problem, the potentials are usually assigned, and the charge distribution becomes the unknown. 
This therefore forms an integral equation, which is the premise for the method of moments capacitance extractor.

\subsection{Multilayer Green's Function in Electrostatics}
\label{sec:multilayer_green_function}
In this section, we discuss a general method to compute the multilayer Green's function using Hankel transforms~\cite{Chew1995,Li2021}.
We will show that the Green's function of a multilayer structure is governed by the Hankel transform~\cite{Michalski1997}, which can be reduced to the two-layer form~\eqref{two_layer_green_func}.
Moreover, a set of numerical procedures is provided for solving the surface integral equation~\eqref{integral_form} efficiently and robustly.

\begin{figure}[t]
  \centering
  \includegraphics[width=\linewidth]{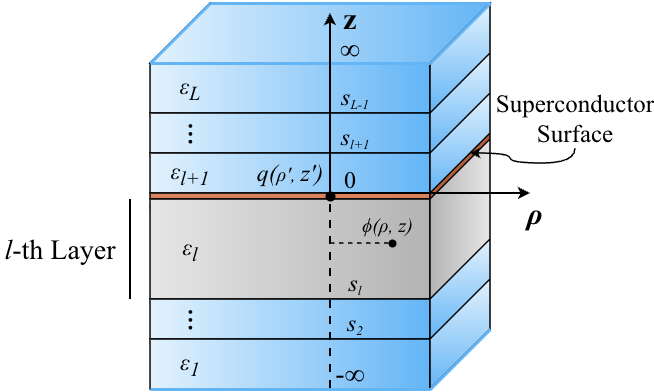}
  \caption{Schematic of an $L$-layer structure defined in cylindrical coordinates $(\rho, z)$. The system consists of multiple regions with varying permittivities (from $\epsilon_1$ to $\epsilon_L$), bounded by a semi-infinite top layer and bottom layer. A superconductor surface is situated at the interface $z = 0$, where a source point charge $q(\rho^\prime,z^\prime)$ is placed. The electrostatic potential $\phi(\rho, z)$ can be calculated at any location in the system, depicted here as being inside the $l$-th layer.} 
  \label{figure:multilayer}
\end{figure}

A real quantum device (e.g., flip-chip) is a stacked planar structure, as depicted in Fig.~\ref{figure:multilayer}, with the thickness of the superconducting films much smaller than the size of the stackup heights and the planar structures.  
We can then assume the thickness of the films to be zero. 
Therefore, the charges that reside on the superconductor surfaces are considered to be located at the interface between two adjacent dielectric layers.
This assumption motivates us to use the multilayer Green's function to solve this problem. 

The spectral differential equation approximation method (SDEAM) was proposed~\cite{Li2021,Li2022} to derive the multilayer Green's function.
By following a similar procedure, we can compute the Green's function in two steps:
\begin{enumerate}
    \item The Hankel transform is applied to the Poisson equation in the spatial domain, yielding a 1D ordinary differential equation with an analytic solution for the spectral Green's function. To incorporate the singularity extraction method, we partition the spectral Green's function into the primary field contribution and the residual scattered field.
    \item The spatial Green's function of the primary field is recovered analytically via the inverse Hankel transform. Conversely, the spatial term of the residual scattered field lacks a closed-form solution, which requires the numerical evaluation of an infinite integral containing a Bessel function. We compute the numerical integration through the Ogata quadrature method~\cite{Ogata2005}. Unlike conventional algorithms, the Ogata quadrature is specifically designed for integrals involving Bessel functions. It approximates the infinite integral as a discrete summation by utilizing specialized quadrature nodes based on the zeros of the Bessel function. By evaluating the integrand at these optimal sampling points and applying corresponding weights, the method effectively mitigates the quadrature errors.
    % This provides a highly robust and numerically stable solution, ensuring efficient computation of the spatial-domain scattered field.}
\end{enumerate}

In our problem, as in Fig.~\ref{figure:multilayer}, 
each planar medium is assumed to be isotropic. We establish a cylindrical coordinate system with radial distance $\rho$ and axial coordinate $z$; the azimuth $\varphi$ is omitted due to the isotropic nature of each layer. 

In the cylindrical coordinate system, we can rewrite the integral equation in \eqref{integral_form} as:
\begin{equation}
    \int_{\mathcal{S}} d\rho^\prime dz^\prime 
    G(\rho, z;\rho^\prime, z^\prime)q(\rho^\prime,z^\prime) = \phi(\rho,z),
    \label{eq:sie}
\end{equation}
where the integration is performed over the superconducting surfaces $\mathcal{S}$. With the electrostatic Green's function $G(\rho, z;\rho^\prime, z^\prime)$ and the applied potential $\phi(\rho,z)$, the charge distribution $q(\rho^\prime,z^\prime)$ is to be solved.

The source point is located on one of the interfaces at $z^\prime$, with radial distance $\rho^\prime = 0$. We can simplify the notation by reducing $G(\rho, z; \rho^\prime, z^\prime)$ to $G(\rho, z; z^\prime)$ without loss of generality.
The spatial domain Green's function satisfies Poisson's equation with a delta-function source:
\begin{equation}
    \label{eq:pde_green_function}
    \nabla^2 G(\rho,z;z^\prime)=-\frac{1}{\epsilon_0\epsilon_r(z)} \frac{\delta(\rho)}{2\pi \rho} \delta(z-z^\prime).
\end{equation}
We designate the transformation pair using the common notation: $F(\rho) \overset{\mathcal{H}}{\longleftrightarrow} \tilde F(\lambda)$. 
After applying the Hankel transform of the form: 
\begin{equation}
    \label{eq:hankel_transform}
    \tilde F (\lambda) = \mathcal{H}\{F(\rho)\} = \int_{0}^{\infty}F(\rho)J_0(\lambda \rho) \rho d\rho, 
\end{equation}
we obtain the governing ordinary differential equation as
\begin{equation}
    \label{eq:ode_green_function}
    \epsilon_r(z)\frac{d^2}{dz^2}\tilde G(z,\lambda;z^\prime) - \lambda^2 \epsilon_r(z) \tilde G(z,\lambda;z^\prime)=-\frac{\delta(z-z^\prime)}{2\pi \epsilon_0},
\end{equation}
where $\tilde G(z,\lambda; z^\prime)$ is the spectral domain Green's function.

Boundary conditions at the $l$-th dielectric interface between layers $l$ and $l+1$ are described as 
\begin{equation}
\label{eq:boundary_interface}
\begin{cases}
\begin{aligned}
    \displaystyle  \tilde G|_{z_{\text{int}, l}^+} &= \tilde G|_{z_{\text{int}, l}^-}, \\
    \displaystyle   \epsilon_{l + 1} \left.\frac{d \tilde G}{dz}\right|_{z_{\text{int}, l}^+} &= \epsilon_{l} \left.\frac{d \tilde G}{dz}\right|_{z_{\text{int}, l}^-},
\end{aligned}
\end{cases}
\end{equation}
where $\epsilon_{l},\epsilon_{l + 1}$ are the permittivities at $l$-th layer and ($l$+1)-th layer, respectively.
Note that these boundary conditions are derivable from Maxwell's equations.

Additionally, the Green's function vanishes at infinity for the top and bottom dielectric layers, as specified by Coulomb's interaction, shown as,
\begin{equation}
    \label{eq:boundary_infinity}
    \tilde G|_{z=-\infty} = \tilde G|_{z=+\infty} = 0.
\end{equation}

The solution for the spectral domain Green's function in each layer is represented as $\tilde G_l = a_l e^{\lambda z} + b_l e^{-\lambda z}$ with two unknown coefficients $a_l$ and $b_l$. 
By enforcing the $2L$ boundary conditions, we convert the multilayer Green's function problem into a $2L$-dimensional linear system. Solving this linear system determines a unique set of solutions for the $2L$ unknown coefficients. Consequently, $a_l(\epsilon_1, \dots , \epsilon_L, s_2, \dots, s_{L-1})$ and $b_l(\epsilon_1, \dots , \epsilon_L, s_2, \dots, s_{L-1})$ are explicitly obtained as functions of the dielectric permittivity $\epsilon_l$ and height $s_l$ for each layer.

A key advantage of this method is its analytical applicability in handling singularity. When the source and field points are located at the same interface, directly recovering the spatial Green's function using numerical integration is unstable due to the $1/r$ singularity, where $r=\sqrt{(z-z^\prime)^2 + \rho^2}$. To address this, we implement a singularity extraction method by partitioning the total spectral Green's function $\tilde G_l$ into a primary field term $\tilde G^\text{prm}_l$ and a residual scattered field term $\tilde G^\text{sct}_l$: 
% we extract the primary field Green's function $\tilde G^\text{prm}$ from the spectral domain Green's function, while the residual term is referred to as the scattered field Green's function $\tilde G^\text{sct}_l$. 
% The primary field and scattered field functions satisfy
\begin{equation}
\begin{aligned}
    \label{eq:primary_green_function}
    \tilde G^\text{prm}_l &= \frac{1}{4\pi \epsilon_0 \epsilon^{\text{eff}}_l} \frac{e^{-\lambda |z-z^\prime|}}{\lambda},\\
    \tilde G^\text{sct}_l &= \tilde G_l - \tilde G^\text{prm}_l,
\end{aligned}
\end{equation}
where $\epsilon^{\text{eff}}_l$ is the effective dielectric permittivity of the $l$-layer, which satisfies $\epsilon^{\text{eff}}_l = (\epsilon_l + \epsilon_{l+1}) / 2$. 
% We account for the $1/r$ singularity, where $r=\sqrt{(z-z^\prime)^2 + \rho^2}$.

Applying the inverse Hankel transform, we derive the spatial Green's function for the primary field shown as
\begin{equation}
    \label{eq:primary_spatial}
    G^\text{prm}_l = \mathcal{H}^{-1}\{\tilde G^\text{prm}_l \} = \frac{1}{4\pi \epsilon_0 \epsilon^{\text{eff}}_l} \frac{1}{\sqrt{\rho^2 + (z-z^\prime)^2}}.
\end{equation}
The Green's function $G^\text{prm}_l$ mirrors the electrostatic Green's function in the two-layer structure. 

By analytically isolating $\tilde G^\text{prm}_l$, we properly account for the $1/r$ singularity in the spatial domain, ensuring that the residual scattered field $\tilde G^\text{sct}_l$ is straightforward to integrate numerically.
Hence, the Ogata quadrature method can be used to efficiently compute the Hankel transform $\mathcal{H}^{-1}\{ \tilde G^\text{sct}_l \}$, resulting in the total field Green's function as
\begin{equation}
\label{multi_layer_green_function}
    G_l =  G^\text{prm}_l +  G^\text{sct}_l.
\end{equation}

\subsection{Discretization and Capacitance Matrix}\label{sec:cap_matrix}
In the SIE simulation, it is generally required to discretize geometries into surface elements. By applying appropriate boundary conditions (excitations), the capacitance matrix can be extracted.

First, by discretizing the metallic surfaces into triangular elements $\{T_j\}$,  
the surface charge density $q(\v{r}^\prime)$, as the unknown in \eqref{integral_form}, can be written as:
\begin{equation}\label{discretize_basis_function}
  q(\v{r}^\prime) = \sum_{j=1}^N a_j h_j(\v{r}^\prime),
\end{equation}
where $N$ is the total number of triangle elements, 
and $h_j(\v{r}^\prime)$ is the basis function on $T_j$. 
The simplest form of $h_j(\v{r}^\prime)$ is the pulse basis function:
\begin{equation}\label{basis_function}
  h_j(\v{r}^\prime) = 
  \begin{cases}
    \frac{1}{A_j}, & \text{if } \v{r}^\prime \in T_j, \\
    0,              & \text{otherwise},
\end{cases}
\end{equation}
where $A_j$ is the area of $T_j$. The definitions in \eqref{discretize_basis_function} and \eqref{basis_function} 
indicate that $q(\v{r}^\prime)$ is approximated as uniformly distributed over each $T_j$.
Consequently, the variation of $q(\v{r}^\prime)$ is fully captured by the changes in $a_j$ across $T_j$. 
Therefore, the geometries must be finely discretized in regions where the surface charge density varies rapidly, which sets the mesh density requirements for a given problem.

Then the equation \eqref{integral_form} is transformed into a matrix equation using Galerkin method, which yields:
\begin{equation}\label{matrix_eqn}
  \mat{G} \cdot \v{q} = \vg{\phi},
\end{equation}
where the matrix and the vector elements can be expressed as:
\begin{align}
  \left[ \mat{G} \right]_{ij} &= 
  \int_{T_i} d\v{r} h_i(\v{r}) 
  \int_{T_j} d\v{r}^\prime G(\v{r}, \v{r}^\prime) h_j(\v{r}^\prime), \label{gf_element} \\
  \left[ \vg{\phi} \right]_{i} &= 
  \int_{T_i} d\v{r} h_i(\v{r}) \phi(\v{r}), \label{potential_element} \\
  \left[ \v{q} \right]_{j} &= a_j.
\end{align}
In the matrix system, the standard singularity extraction in \eqref{gf_element} and the charge neutrality constraint need to be applied to ensure an accurate simulation of $\v{q}$. The singularity extraction~\cite{Yla-Oijala2003} addresses the singular kernel of $\left[ \mat{G} \right]_{ij}$ when $T_i$ and $T_j$ overlap, and the charge neutrality constraint avoids the arbitrariness of the total charge in the system~\cite{qian2009fast,xia2016enhanced}, see Appendix~\ref{sec:charge_neutrality}. 

Finally, the capacitance matrix can be obtained from the definition of its elements:
\begin{equation}\label{cap_matrix_element}
  \left[\mat{C}\right]_{pq} = \frac{Q_p}{V_q} \bigg|_{ V_{k} = 0 \text{ for } k\neq q},
\end{equation}
where $p$ and $q$ are the indices of the metallic objects. 
By applying the excitations: $V_q = 1$~V and $V_k=0$ for $k\neq q$, one can solve for the charge distribution on metallic objects. By summing up the total charge on object $p$ to get $Q_p$,  $\left[\mat{C}\right]_{pq}$ is obtained. Therefore, extracting the full capacitance matrix involves solving \eqref{matrix_eqn} with multiple right-hand sides.

\subsection{Surface Dielectric Participation Ratio}\label{sec:surf_pr}
After the charge distribution $\v{q}$ is solved in \eqref{matrix_eqn},
the electric field at an arbitrary point $\v{r}$ can be computed as:
\begin{equation}\label{post_e_field}
  \v{E}(\v{r}) = - \nabla \phi(\v{r})
  = - \sum_{j=1}^N a_j \int_\mathcal{S} d\v{r}^\prime \nabla G(\v{r}, \v{r}^\prime) h_j(\v{r}^\prime).
\end{equation}
Then the total energy within a volume $\Omega$ is:
\begin{equation}\label{post_energy}
  U = \frac{1}{2} \int_\Omega d\v{r} \epsilon(\v{r}) |\v{E}(\v{r})|^2.
\end{equation}
\begin{figure}[t]
  \centering
  \includegraphics[width=\linewidth]{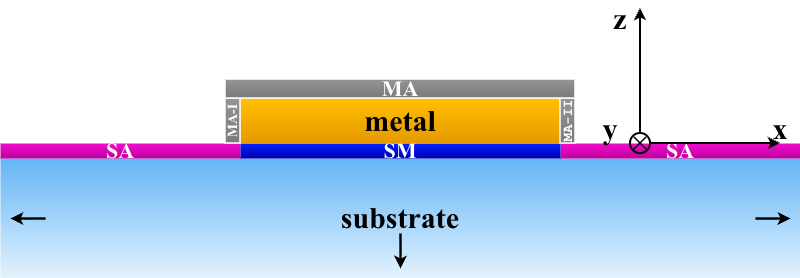}
  \caption{The cross-section of the dielectric loss interfaces. The structure consists of a substrate-metal interface (SM) sandwiched between a substrate and a superconducting metal conductor. The metal conductor is encapsulated by metal-air interfaces (MA, MA-I, and MA-II), while the surrounding regions are covered by substrate-air interfaces (SA). The coordinate system indicates the $x$, $y$, and $z$ axes, with the $y$-axis oriented into the plane.} 
  \label{figure:interfaces}
\end{figure}

In the superconducting quantum circuit (SQC), the relaxation of superconducting qubits and resonators 
through dielectric losses is attributed to the coupling between the electromagnetic qubit mode and the two-level 
system (TLS)~\cite{Martinis2005}, particularly at the interfaces of superconductors and dielectrics. 
The relaxation rate $\Gamma$ at zero temperature due to the dielectric losses at these interfaces is expressed as:
\begin{equation}
  \Gamma = \omega_q \sum_i P_i \tan \delta_i,
\end{equation}
where $\omega_q$ is the qubit frequency, $\tan \delta_i$ characterizes the intrinsic dielectric loss of the $i$-th interface, and $P_i$ is the EPR, defined as the fraction of the electric energy within the lossy dielectric volume $\Omega_i$ 
to the total energy in the entire volume $\Omega$:
\begin{equation}\label{par_def}
  P_i =\frac{\int_{\Omega_i} d\v{r} \epsilon(\v{r}) |\v{E}(\v{r})|^2}{\int_{\Omega} d\v{r} \epsilon(\v{r}) |\v{E}(\v{r})|^2},
\end{equation}
where the total energy in the denominator can be calculated with the charges $Q_k$ and assigned potentials $V_k$ of all superconductors without volume integration:  
\begin{equation} \label{eq:energy_total}
    \int_{\Omega} d\v{r} \epsilon(\v{r}) |\v{E}(\v{r})|^2 = \sum_k Q_k V_k .
\end{equation}
The energy in $\Omega_i$ then needs to be computed numerically for arbitrary structures. 
The lossy dielectric interfaces of interest are shown in Fig.~\ref{figure:interfaces}, 
including substrate-metal (SM), substrate-air (SA), and metal-air (MA) interfaces. 

Noting that the thickness of the metal is usually much smaller than the size of the structure,
the electric field energy within the volume of MA-I and MA-II in Fig.~\ref{figure:interfaces} 
contributes less than $2\%$ of the total contribution from the MA volume~\cite{Murray2018}. 
Therefore, we can simplify the evaluation of the EPR by ignoring the contributions from these 
two interfaces and approximating the thickness of the metal as zero. 

\begin{figure}[t]
  \centering
  \includegraphics[width=\linewidth]{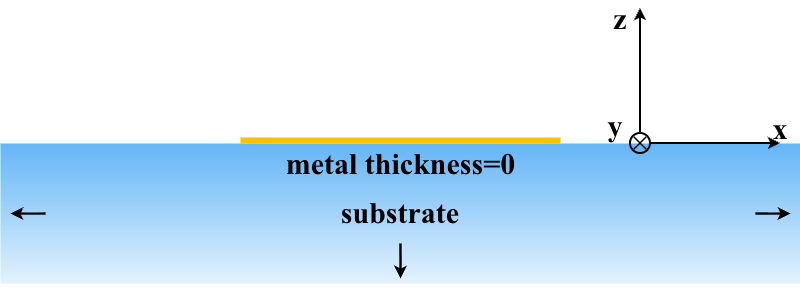}
  \caption{The simplified cross-section for the simulation. The thickness of the metal is approximated as zero. The dielectric constants of the SM and SA interfaces are assumed to be the same as the substrate, while the MA interface is assumed to be air.} 
  \label{figure:interfaces_simplify}
\end{figure}
To further simplify the calculations to be treated with SIE, 
we treat the substrate layer as infinitely thick and wide,  
i.e., the electromagnetic field underneath the substrate is infinitesimally small, and 
the field is well confined near the superconductor. 
This assumption becomes valid if the gaps between the conductors are much smaller than the substrate thickness, 
and the strong-field regions are far from the boundaries. 
To avoid over-complicating the Green's function due to the very thin ($\sim$ nm) dielectric interfaces, 
the dielectric constants of the SM and SA interfaces are assumed to be the same as the substrate,
while the MA interface is assumed to be air. 
The simplified cross-section is shown in Fig.~\ref{figure:interfaces_simplify}.
Although the dielectric constants of these interfaces are ignored, 
we can take them into account when computing the electrical energy within the volumes 
by matching the boundary conditions \cite{Murray2018}. 

With these assumptions, we treat the problem in the electrostatic limit.
Directly plugging \eqref{multi_layer_green_function} into \eqref{gf_element} 
yields the matrix equation \eqref{matrix_eqn} to be solved. 

\subsection{Boundary Singularities}\label{sec:sig_mesh_refine}
At the edges and corners of the conductors, 
the charge density becomes singular under the electrostatic excitations~\cite{Meixner1972}. 
This makes it difficult to calculate the charges and field accurately using numerical methods. 
One approach to mitigating this accuracy issue is to apply the $h$-refinement near the boundary, 
i.e., discretizing the boundary region with dense meshes, to approximate the singular charge density. 
However, it may not always be feasible to generate a conformal mesh with a large size gradient 
between adjacent triangle elements.
Meanwhile, the dramatic increase in the number of triangle elements makes it computationally costly.
Fortunately, the basis function, as defined in \eqref{basis_function}, does not necessarily require the mesh to be conformal. 
Hence, we can start from a coarse mesh, 
and apply a few techniques to refine the mesh near the edges with an affordable increase in the number of unknowns. 

\begin{figure}[t]
  \centering
  \includegraphics[width=\linewidth]{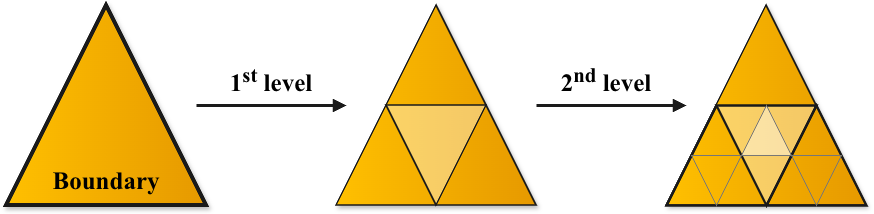}
  \caption{Homogeneous mesh refinement on a boundary triangle. 
  Each triangle is refined into four congruent triangles,
  where the extra three vertices reside at the center of the edges. 
  Multiple levels of refinement can be applied. 
  } 
  \label{figure:refine1}
\end{figure}
The first technique is the homogeneous mesh refinement, which breaks a boundary triangle into four smaller ones, as depicted in Fig.~\ref{figure:refine1}.
Multiple levels of refinement can be adopted using this scheme.
The generated mesh is non-conformal since only the boundary triangles are refined. 
An apparent advantage of this technique is that 
the size of the triangle element reduces exponentially as it gets closer to the boundary, 
which is favorable to approximately capture the increment of charge density towards the boundary. 
Meanwhile, the mesh quality is well controlled as the aspect ratio of the refined triangles is the same as the original. 
However, it is impractical to apply this refinement to many levels: 
the number of triangle elements increases exponentially with the number of refinements, thereby significantly increasing the computational cost. 

\begin{figure}[t]
  \centering
  \includegraphics[width=0.78\linewidth]{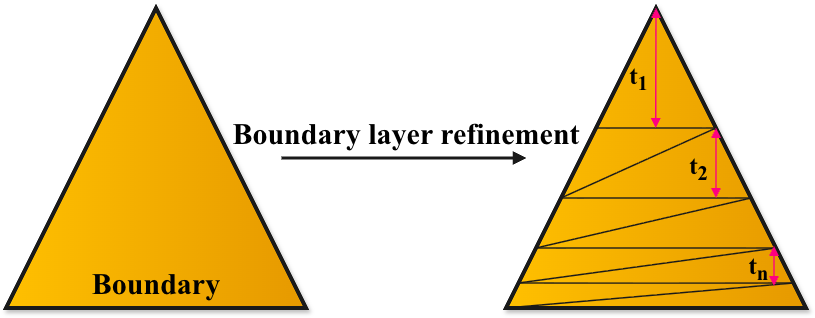}
  \caption{Boundary layer refinement of a boundary triangle.
  The height of each refined triangle $t_i$ decreases exponentially to 
  capture the singularity accurately.} 
  \label{figure:refine2}
\end{figure}

A remedy to avoid the exponential increase in the number of triangle elements
is to utilize a boundary layer mesh. As depicted in Fig.~\ref{figure:refine2}, 
only the boundary layer triangles are refined into a stack of small triangles with a height of $t_n$. 
A constant ratio $t_{n-1} / t_n$ can be easily assigned so that the triangle size reduces exponentially towards the boundary
while the number of triangles increases linearly.
One of the drawbacks of this technique is that the aspect ratio is not maintained.

With the above two techniques, we can first apply the homogeneous mesh refinement with only a few levels to a coarse mesh.
The mesh sizes decrease exponentially, 
and the charge density variations along both lateral and normal directions can be well approximated. 
After that, a boundary layer refinement can be applied to the outermost boundary layer to aggressively refine the mesh along the normal direction. 
This refinement is suitable for problems with conductors on a planar surface, 
where the charge density varies rapidly along the normal direction of the boundary, while the lateral variations are much slower. 

This treatment of the boundary mesh can be regarded as a local mesh refinement method:
the discretization is significantly improved in the region where the unknowns vary rapidly. 
The effectiveness of this method will be shown later in Section~\ref{sec:results}.

\subsection{Energy Integration}
With \eqref{post_e_field}, \eqref{post_energy} and \eqref{par_def},  
one can evaluate the electric energy $U_i$ in $\Omega_i$ in order to find the EPR.
Assuming the thickness of the contaminated dielectrics is $\delta$,
\eqref{post_energy} can be written as:
\begin{equation}\label{energy_integral}
  U_i = \int_0^\delta u_i(z) dz,
\end{equation}
where $u_i(z)$ is the surface energy density:
\begin{equation}\label{integral_form_of_energy}
  u_i(z) = \frac{1}{2} \int dx \int dy \epsilon (x, y, z) |\v{E}(x, y, z)|^2.
\end{equation}
From the boundary conditions of the electric field: at $z=0$, where the conductor resides, 
the surface charge density is proportional to the normal electric field, and the tangential electric field vanishes. 
Therefore, the integral
\begin{equation}\label{charge_boundary_condition}
  \int dx \int dy E_z(x, y, 0) \propto Q_{\text{total}},
\end{equation}
which converges. However, the electric field is singular at the boundary of the conductor, which results in 
a divergent $u_i(z=0)$ due to the square of the electric field. 
Therefore, regardless of the thickness $\delta$, 
it is not valid to approximate the energy integral in \eqref{energy_integral} as $\delta u_i(z=0)$. 

An appropriate calculation of the integral in \eqref{energy_integral} should avoid the evaluation at the singular point at $z=0$, 
while accurately capturing the converged integral. Using the Legendre-Gauss quadrature~\cite{Abramowitz1964}, 
$u_i(z)$ is evaluated only at a set of roots of the $n$-th Legendre polynomial $z_k$ where $0 < z_k < \delta$, such that:
\begin{equation}
   U_i \approx \sum_k w_k u_i(z_k),
\end{equation}
with $w_k$ being the weighting factor at the quadrature point $z_k$.

\section{Numerical Results}\label{sec:results}
In this section, we present a comprehensive validation of the proposed method using two distinct test cases: a double-layer (2-L) structure and a triple-layer (3-L) structure.

For the 2-L case, we employ a coplanar capacitor (CPC) positioned at the interface between an air layer and a substrate layer. In this configuration, the Green's function satisfies a simple form, $G\sim 1/|\mathbf{r}-\mathbf{r}'|$, 
% on both sides of the interface, 
where the electric field induced by the source points scales as $\mathbf{E}\sim\mathbf{r}/|\mathbf{r}|^3$. However, both the Green's function and the electric field exhibit singularities at the metal edges. Consequently, the 2-L structure serves as an ideal platform to validate our singularity extraction method and numerical robustness.

For the 3-L case, we construct a grounded coplanar waveguide (GCPW) model. In this setup, the conductors are deposited on a central substrate, while the upper and lower semi-infinite regions are filled with air. 
This model rigorously tests the efficiency and accuracy of the EM simulator, as the electric field distribution becomes complex.
% within the gaps between interfaces. 
Moreover, the multilayer Green's function must be employed to account for the effects arising from the layered structure.
Since no closed-form solution exists for the multilayer Green's function in the spatial domain, we employ various numerical techniques to accelerate the calculation, including 
% series expansion, 
Ogata quadrature, lookup tables, etc. 
The 3-L example serves to validate the efficiency of the multilayer Green's function implementation.

To establish a reference for the numerical method, we derive the analytical capacitance and the EPR using the conformal mapping method. Furthermore, the capacitance and the EPR are calculated via the Finite Element Method (FEM) using ANSYS Maxwell~\cite{AnsysMaxwell} to facilitate a fair comparison between our proposed method (SesQ) and a general-purpose electrostatic simulation suite.

We demonstrate that both the capacitance and the EPR can be calculated accurately and efficiently using the SIE method. Notably, the accuracy of the EPR calculation is significantly improved compared to FEM simulations using ANSYS Maxwell. 
Subsequently, this method is applied to the design of floating qubit patterns to achieve reduced EPRs.

All simulations were performed on a workstation equipped with an Intel Xeon Platinum 8580 CPU @ 2.50 GHz (60 cores) and 384 GB memory. FEM simulations in this section were performed using ANSYS Electronics Desktop 2025 R1.

\subsection{Validation by CPC}
\begin{figure}[t]
    \centering 
    \includegraphics[width=\linewidth]{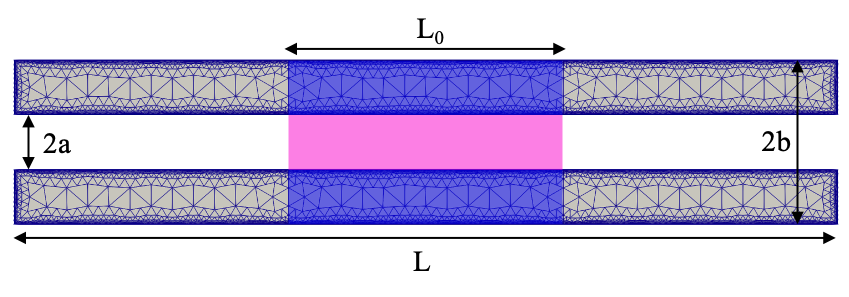} 
    \caption{A coplanar capacitor with the geometry denoted by $a$, $b$ and $L$.
    The central region with length $L_0$ consists of 
    the SM interface (blue) and the SA interface (pink).
    Triangles showing the non-uniform mesh discretization.
    }\label{figure:psm_psa_analytical_sch} 
\end{figure}
To numerically validate the proposed method, we select a Coplanar Capacitor (CPC) as a benchmark, as the analytical solution for its electric field can be derived using the conformal mapping technique~\cite{Murray2018}.
Although the analytical solution strictly applies to an infinitely long CPC, a sufficiently long finite structure yields an electric field at the center $L_0$ that effectively matches the analytical solution, as illustrated in Fig.~\ref{figure:psm_psa_analytical_sch}. 
We term this strategy the de-embedding technique, which suppresses the end effects arising from the finite truncation of the simulation boundaries. It is important to note that while only the central region is selected for field evaluation, the entire length $L$ must be included as source points to ensure high numerical accuracy.

Then the solved charges on the conductor pads can be used to extract the capacitance and the EPR, as discussed
in Sections~\ref{sec:cap_matrix} and \ref{sec:surf_pr}.
The analytical expression of the capacitance~\cite{Garg2013}, contributed by the electric field in both the air and dielectric regions, takes the form:
\begin{equation}\label{analytical_c}
  C_{\text{cpc}} = \frac{1}{2} \epsilon_0 (\epsilon_{\text{sub}}+ 1) \frac{K(k^\prime)}{K(k)},
\end{equation}
where $K(k)$ is the complete elliptic integral of the first kind~\footnote[1]{Note that in most numerical packages,  
the complete elliptic integral of the first kind is implemented as $K(m)$ with $m=a^2/b^2$.}, 
with the argument $k={a}/{b}$ and complementary argument $k'=\sqrt{1-k^2}$.
By assuming thin metal thickness, the EPR at the SM interface can be found~\cite{Murray2018}:
\begin{align}\label{analytical_psm}
    \begin{split}
    P_\text{SM}\left(\frac{\delta}{a}\right) \approx  &
    \frac{\delta}{a} \frac{\epsilon_\text{sub}^2}{\epsilon_c(\epsilon_\text{sub} + 1)}
    \frac{1}{2(1-k)K(k^\prime) K(k)} \\
    & \cdot \left[ 
      \ln \frac{4a(1-k)}{\delta(1+k)} - \frac{k \ln k}{1+k} +1
      \right],
  \end{split}
\end{align}
where $\delta$ and $\epsilon_c$ are the thickness and the dielectric constant of the contamination interface.
Since in Fig.~\ref{figure:psm_psa_analytical_sch}, only the central slot between the conductors 
is taken into account as the SA interface, 
the EPR at SA interface should be evaluated 
by matching the boundary conditions~\cite{Wenner2011,Murray2018,Murray2020,Sandberg2012}.

By applying the FEM/SIE methods and the analytical expressions, we simulate CPCs on a silicon substrate ($\epsilon_{\text{sub}}$ = 11.9)
with $L=800$ $\mu$m and $L_0=100$ $\mu$m for a few combinations of $a$ and $b$. 
Different mesh densities are applied to each of the combinations to study the convergence. 
In this study, we quantify the cost of the numerical methods by the simulation wall time.

\begin{figure}[h]
    \centering 
    \includegraphics[width=\linewidth]{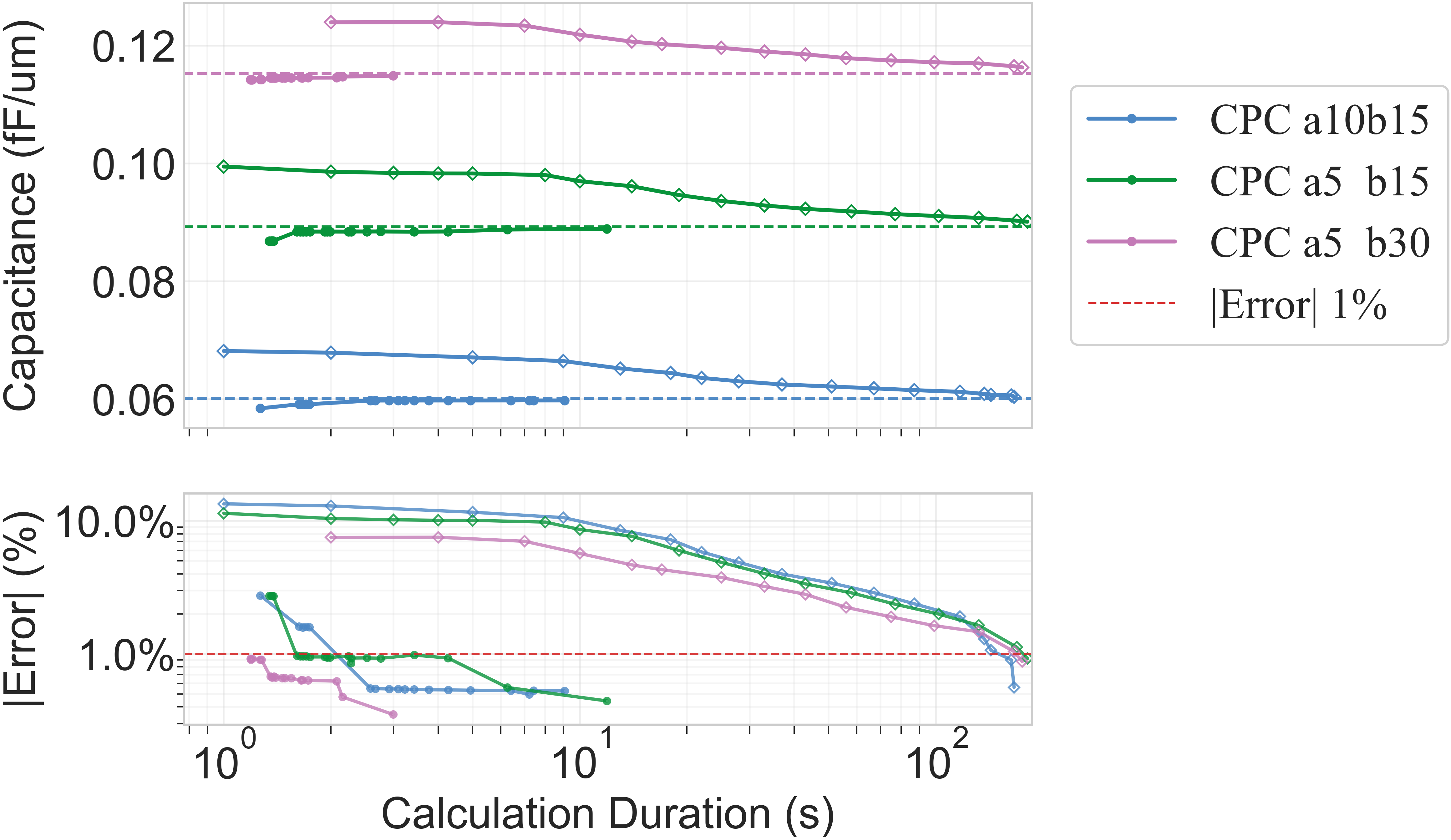} 
    \caption{The convergence of the capacitance simulated by the FEM method (diamonds) and the proposed SIE method (circles) to the analytical values (dotted lines).
    The CPCs are deposited on a silicon substrate ($\epsilon_{\text{sub}}$ = 11.9) for three geometric configurations: $(a, b) = (10, 15)\ \mu\text{m}$ (blue), $(5, 15)\ \mu\text{m}$ (green), and $(5, 30)\ \mu\text{m}$ (purple).
    The red dotted lines indicate a $1\%$ absolute relative error boundary. 
    The FEM method achieves an accuracy of 1\% in approximately 160 s, whereas the SIE method requires roughly 3 s.
    } 
   \label{figure:cpc_cap} 
\end{figure}

As shown in Fig.~\ref{figure:cpc_cap}, the capacitance converges rapidly with simulation time
with the proposed SIE method and moderately with the FEM method. 
This is as expected, since the Galerkin method is of second order accuracy due to its variational form~\cite{Richmond1991,Harrington1993}.
Note that, for the FEM results, we only take the last adaptive simulation time 
into account for a fair comparison. The actual FEM total time cost, including the adaptive meshing process and iterative simulation steps, 
is roughly three times of the recorded simulation time. 

\begin{figure}[h]
    \centering
    \includegraphics[width=\linewidth]{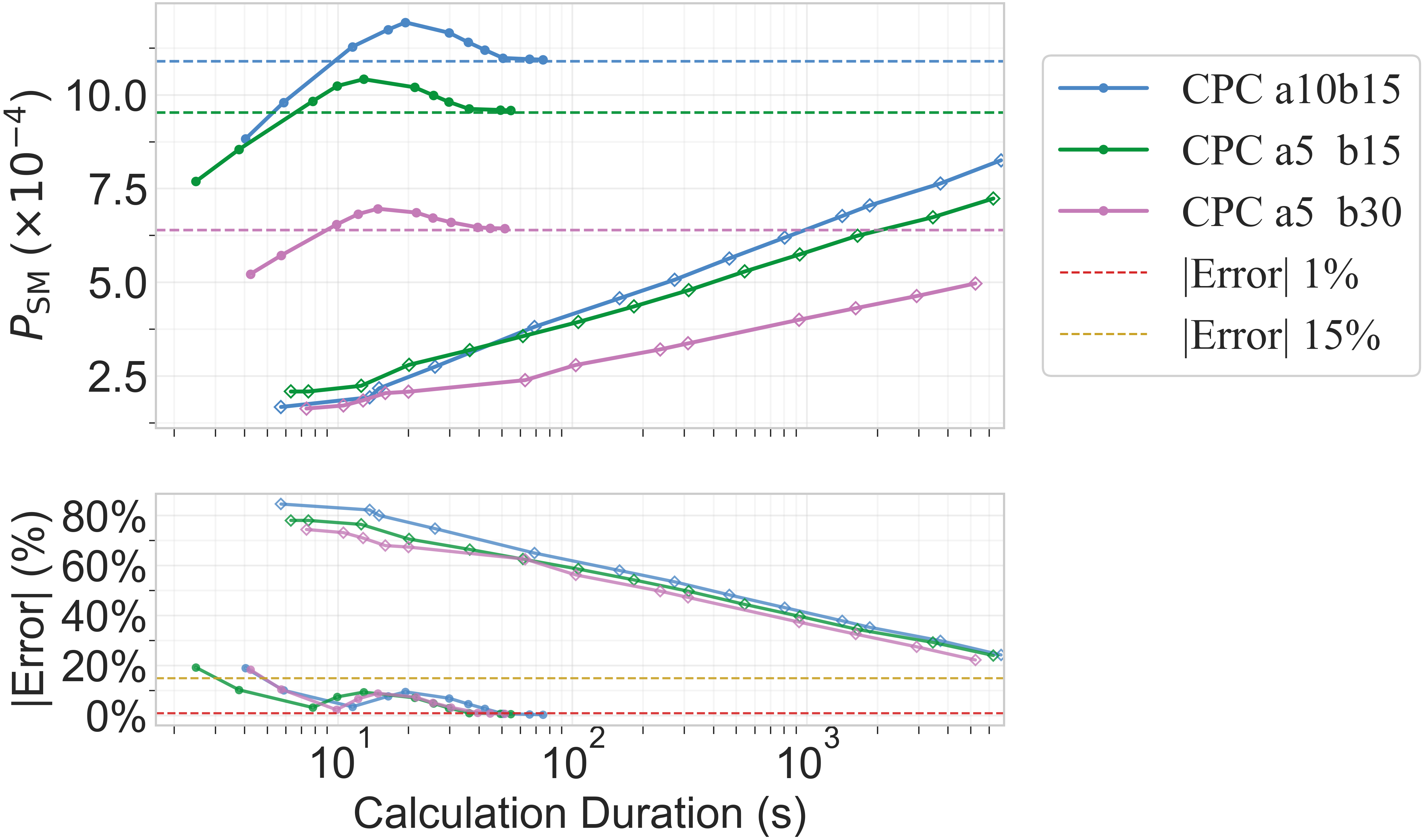} 
    \caption{Comparison of EPR convergence at the SM interface using FEM and SIE. The simulated convergence behaviors of the FEM (diamonds) and SIE (circles) are validated against analytical values (dotted lines) for CPCs on a silicon substrate ($\epsilon_{\text{sub}} = 11.9$). A contamination layer ($\epsilon_c = 11.9$) with a thickness of $\delta = 3$ nm is evaluated across three geometric configurations: $(a, b) = (10, 15)\ \mu\text{m}$ (blue), $(5, 15)\ \mu\text{m}$ (green), and $(5, 30)\ \mu\text{m}$ (purple). 
    The red dotted line indicates a 1\% relative error boundary, and the yellow dotted line indicates a 15\% relative error boundary.
    Notably, the FEM fails to converge due to memory exhaustion after approximately 6700~s with an accuracy of about $15\%$, whereas the SIE achieves the same accuracy in roughly 30~s,
    demonstrating a computational speedup of two orders of magnitude.}
    \label{fig:combined_psm}
\end{figure}
However, achieving good convergence for the EPR requires more effort.
We apply the electrostatic FEM simulator to such a configuration: 
A CPC with $L=800$ $\mu$m, $L_0=100$ $\mu$m and 
the contamination interfaces of $\delta=3$ nm.
Both the substrate and the contamination dielectric constant are assumed to be 
$\epsilon_{\text{sub}} = \epsilon_c = 11.9$ without loss of generality. 

The ultra-thin contamination interfaces are meshed with fine tetrahedrons in order to obtain an accurate 
electric energy within the volumes. 
At the same time, however, the number of unknowns grows significantly due to the multiscale nature of the problem, 
which prevents further mesh refinement at the interfaces. 
As a result, the computational cost becomes an obvious bottleneck. Not only is the time cost high, but the memory cost is also huge, while our workstation has insufficient memory to obtain converged results.

As shown in Fig.~\ref{fig:combined_psm}, the FEM simulation results converge slowly to the analytical values, and
it takes roughly an hour to achieve $15\%$ accuracy. Again, only the last adaptive simulation time is counted. 

The same example, as a comparison, is simulated with the proposed SIE method with non-conforming mesh refinements. 
An initial two-level homogeneous refinement, as in Fig.~\ref{figure:refine1}, 
is first applied to the boundary triangles. 
Then, boundary layer mesh refinement, as in Fig.~\ref{figure:refine2}, is adopted with varying $t_N$.
The resultant mesh is densely meshed near the boundary while the number of unknowns increases moderately.
In Fig.~\ref{fig:combined_psm}, we also show the convergence of the SIE results. 
The simulation time covers all the time consumption to get the EPRs at the SM interface. 
By comparing FEM results with SIE results,
we can conclude that, to get solutions with an accuracy of $15\%$, 
roughly 200 times acceleration can be achieved with the proposed method. 

\subsection{Validation by Grounded CPW}
Although the 2-L CPC example validates the efficiency and accuracy of the proposed SIE method, the Green's function in the 2-L case simplifies to the $1/r$ form. 
Therefore, a 3-L model with analytical solutions for both capacitance and participation ratio is essential to rigorously validate the correctness of the multilayer Green's function and the corresponding numerical approaches.
\begin{figure}[h]    
  \centering 
    \includegraphics[width=0.9\linewidth]{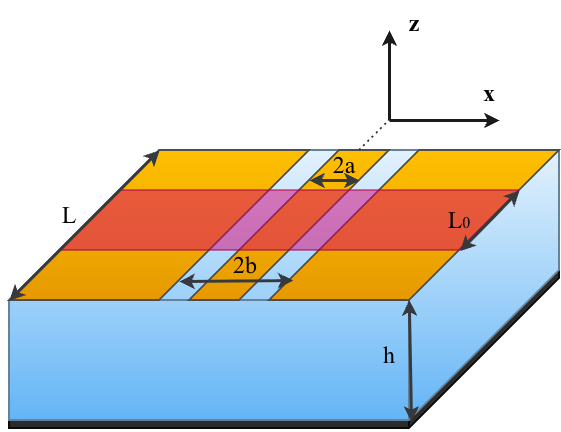} 
    \caption{3D view of a grounded coplanar waveguide (GCPW) geometry. The structure consists of a central signal line separated by gaps from two semi-infinite ground conductor planes on the top surface of a dielectric substrate, with an additional continuous ground conductor plane on the bottom surface. The Cartesian coordinate system indicates the transverse ($x$) and normal ($z$) directions.
    The central region (pink) with length of $L_0$ consists of 
    the superconducting surface (yellow) and the substrate (blue).
    }
    \label{figure:grounded_cpw_sketch} 
\end{figure} 
A Grounded Coplanar Waveguide (GCPW) is selected as the benchmark example, as its analytical solutions for capacitance and electric field can be derived using the conformal mapping technique. Since these analytical solutions apply to infinitely long structures, we introduce a de-embedding technique to ensure that the numerical solutions from finite-length (truncated) models converge to the analytical results of the infinite case. The GCPW model is shown in Fig.~\ref{figure:grounded_cpw_sketch}.

The specific parameters of the layer stack are used to construct a Green's function database, following the methodology discussed in Section~\ref{sec:multilayer_green_function} and Appendix~\ref{sec:prepare_database}. Consequently, all medium properties and boundary conditions are encapsulated within the Green's function. The remaining procedures—such as discretization, enforcement of charge neutrality, and the solution of the matrix equation—remain identical to those in the 2-L case.

Once the surface charges on the conductor are solved, they can be used to compute the capacitance. The analytical expression for the total capacitance is provided by Ghione~\cite{Ghione1983} and takes the following form:
\begin{equation}
\label{analytical_gcpw_c}
C_\text{gcpw} = 2\epsilon_0\epsilon_\text{out} \frac{K(k)}{K(k')} + 2\epsilon_0 \epsilon_\text{in} \frac{K(k_1)}{K(k'_1)},
\end{equation}
where the geometric arguments are defined as $k = a/b$ and $k_1 = \tanh[{\pi a}/({2h})] / \tanh[{\pi b}/({2h})]$. Here, $h$ denotes the thickness of the intermediate substrate (with permittivity $\epsilon_\text{in}$), which is sandwiched between the signal/ground lines and the bottom ground plane. 
The top and bottom mediums outside the intermediate layer extend to infinity with the same permittivity $\epsilon_\text{out}$ ($=1$ if surrounded by air).

Effectively, this structure can be modeled as two capacitors in parallel~\cite{Hanna1985,Li2018}, representing the contributions from the upper air half-space and the intermediate substrate layer, respectively.

Furthermore, we employ the SIE to compute the electric field, incorporating the multilayer Green's function formulation. 
For validation, the electric field is also derived using conformal mapping. 
It is important to note that the analytical expressions differ depending on whether the field point is located within or outside the intermediate layer~\cite{Gillick1993}.

In the region exterior to the intermediate layer (i.e., the upper semi-infinite air space), the electric field takes the form: 
\begin{equation}
\label{gcpw_analytical_electric_field_out}
    E_{\text{out}}(\zeta) =\frac{\phi_0 \cdot b}{K(k')}\frac{1}{\sqrt{[\zeta^2-a^2][\zeta^2-b^2]}},
\end{equation}
where $\phi_0$ is the scalar potential of the signal line, and the complex plane is defined as $\zeta=x+iz$.

Conversely, within the intermediate layer, the electric field is given by: 
\begin{equation}
\label{gcpw_analytical_electric_field_in}
\begin{split} 
E&_{\text{in}}(\zeta) = \frac{\phi_0 \pi \cosh(\frac{\pi a}{2h})\sinh(\frac{\pi b}{2h})}{2h K(k_1')} \\ &\cdot \frac{1}{\sqrt{[\cosh^2(\frac{\pi \zeta}{2 h})-\cosh^2(\frac{\pi a}{2 h})] [\cosh^2(\frac{\pi \zeta}{2 h})-\cosh^2(\frac{\pi b}{2 h})]}}.
\end{split} 
\end{equation}

We focus on the EPR at the SM interface, where the electric field satisfies Eq.~\eqref{gcpw_analytical_electric_field_in}. 
% Due to the complexity of the hyperbolic cosine functions involved, deriving an exact closed-form analytical solution analogous to Eq.~\eqref{analytical_psm} is challenging.
Hence, we derive a closed-form EPR solution for the grounded CPW structure (see Appendix~\ref{sec:psm_intermediate_layer_derivation} for details), which is expressed as:
\begin{equation}
\label{eq:gcpw_psm_closed_form}
\begin{aligned}
\mathcal{L}_a &= \frac{4h e}{\pi} \sinh\left(\frac{\pi a}{h}\right) \frac{1-k_1}{1+k_1},\\
\mathcal{L}_b &= \frac{4h e}{\pi} \exp\left({-\frac{\pi b}{h}}\right) \sinh\left(\frac{\pi b}{h}\right) \frac{1-k_1}{1+k_1},\\
P_\text{SM}(\delta) &\approx \frac{\epsilon_0\epsilon_\text{in}^2 \delta}{\epsilon_c C_\text{gcpw}} \cdot \frac{\pi}{h [k_1' K(k_1')]^2} \left[ \frac{1}{\sinh(\frac{\pi a}{h})} \left( \ln\frac{\mathcal{L}_a}{\delta}  + \frac{\pi a}{h} \right) \right.\\
&\qquad \left.+\frac{1}{\sinh(\frac{\pi b}{h})} \ln\frac{\mathcal{L}_b}{\delta} \right],\\
\end{aligned}
\end{equation}
where $e$ is Euler's number, and $\delta$ and $\epsilon_c$ denote the thickness and dielectric constant of the contamination interface, respectively. Eq.~\eqref{eq:gcpw_psm_closed_form} provides a fast and robust method to compute EPR.

The formula is validated through a semi-analytical approach utilizing numerical integration via QUADPACK~\cite{Piessens1983}. The comparison of the EPR between the semi-analytical method and the closed-form is shown in Table~\ref{tab:psm_comparison}.
\begin{table}[h]
\centering
\caption{Comparison of the participation ratio ($P_\text{SM}$) calculated by the semi-analytical method and the closed-form formula ($a=5\ \mu\text{m}$, $b=30\ \mu\text{m}$, $\delta=3\times 10^{-3}\ \mu\text{m}$, $\epsilon_\text{sub}=11.9$, unit: $\times 10^{-4}$).}
\label{tab:psm_comparison}
\setlength{\tabcolsep}{9pt}
\begin{tabular}{l|c|c|c|c}
\hline
\multirow{2}{*}{Method} & \multicolumn{4}{c}{$h\ (\mu\text{m})$} \\
\cline{2-5}
 & 25 & 35 & 45 & 100 \\
\hline
Semi-Analytical & 7.15503 & 6.71921 & 6.55110 & 6.40298 \\
Closed-Form     & 7.15514 & 6.71930 & 6.55118 & 6.40305 \\
\hline
Abs. Diff.      & 1.09e-4 & 8.83e-5 & 7.87e-5 & 6.62e-5 \\
Rel. Diff. (\%) & 1.52e-3 & 1.31e-3 & 1.20e-3 & 1.03e-3 \\
\hline
\end{tabular}
\end{table}

Utilizing the FEM/SIE methods and the derived analytical approach, we simulate grounded CPWs on a silicon substrate ($\epsilon_{\text{sub}} = 11.9$) surrounded by air. The simulations are conducted with a model length of $L=800$ $\mu$m and a central region length $L_0=100$ $\mu$m across various combinations of geometric parameters $a$, $b$, and $h$. 
To investigate convergence behavior, varying mesh densities are applied to each configuration.

\begin{figure}[h]
    \centering 
    \includegraphics[width=\linewidth]{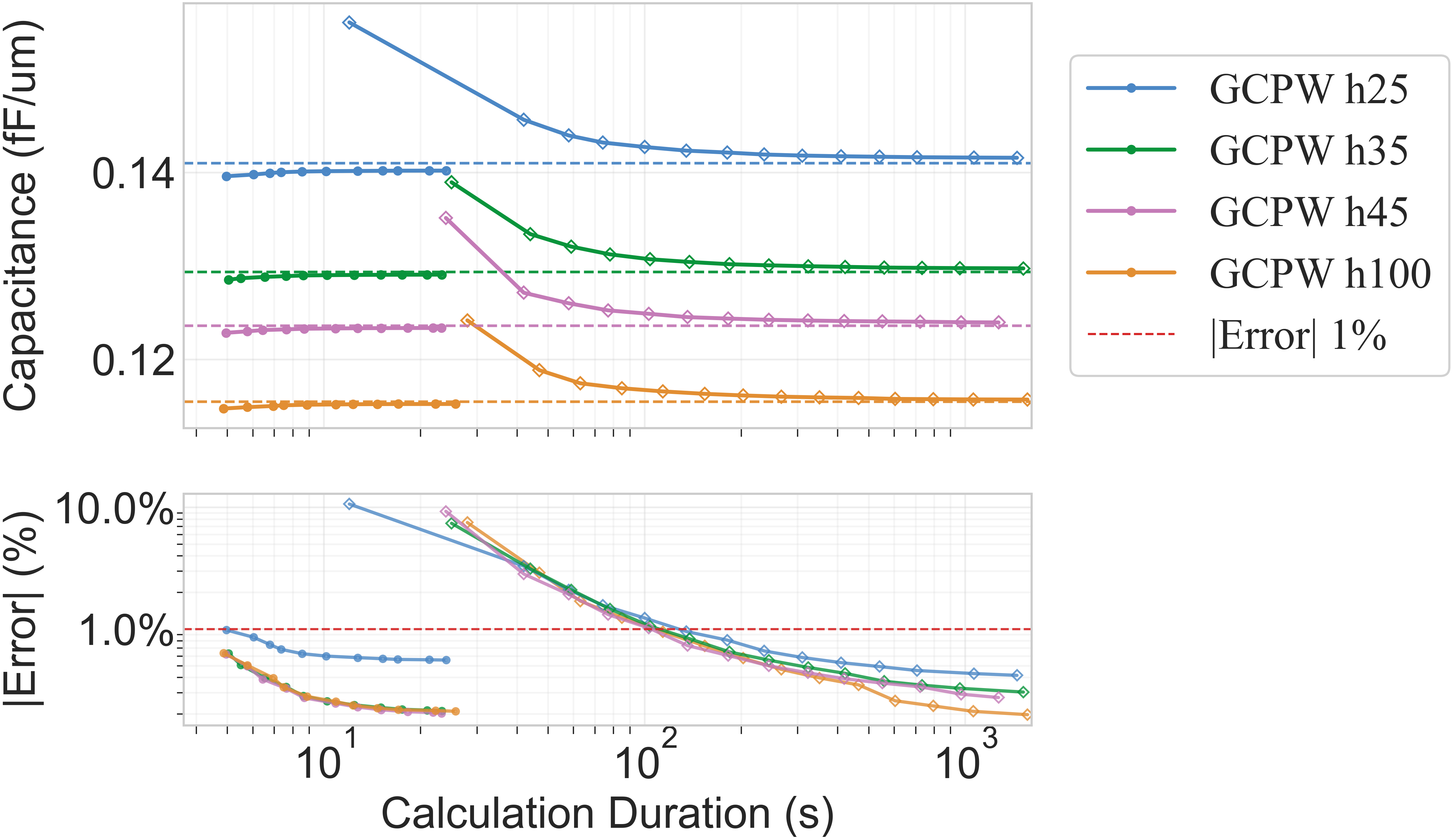} 
    \caption{The convergence of the capacitance simulated by the FEM method (diamonds) and 
    the proposed SIE method (circles) to the analytical values (dotted lines). The GCPWs are deposited on a silicon substrate ($\epsilon_{\text{sub}}$ = 11.9) for $a=5\ \mu$m, $b=30\ \mu$m and four intermediate layer thicknesses: $h=25\ \mu$m (blue), $h=35\ \mu$m (green), $h=45\ \mu$m (purple), $h=100\ \mu$m (orange).
    The red dotted line indicates a 1\% relative error boundary. 
    The FEM method achieves 1\% accuracy in approximately 130 s, whereas the SIE method requires roughly 10 s.
    } 
   \label{figure:gcpw_cap} 
\end{figure}

As illustrated in Fig.~\ref{figure:gcpw_cap}, the capacitance calculated using the proposed SIE method converges rapidly with respect to simulation time, whereas the FEM method exhibits a more moderate convergence rate.

\begin{figure}[htbp]
    \centering
    \includegraphics[width=\linewidth]{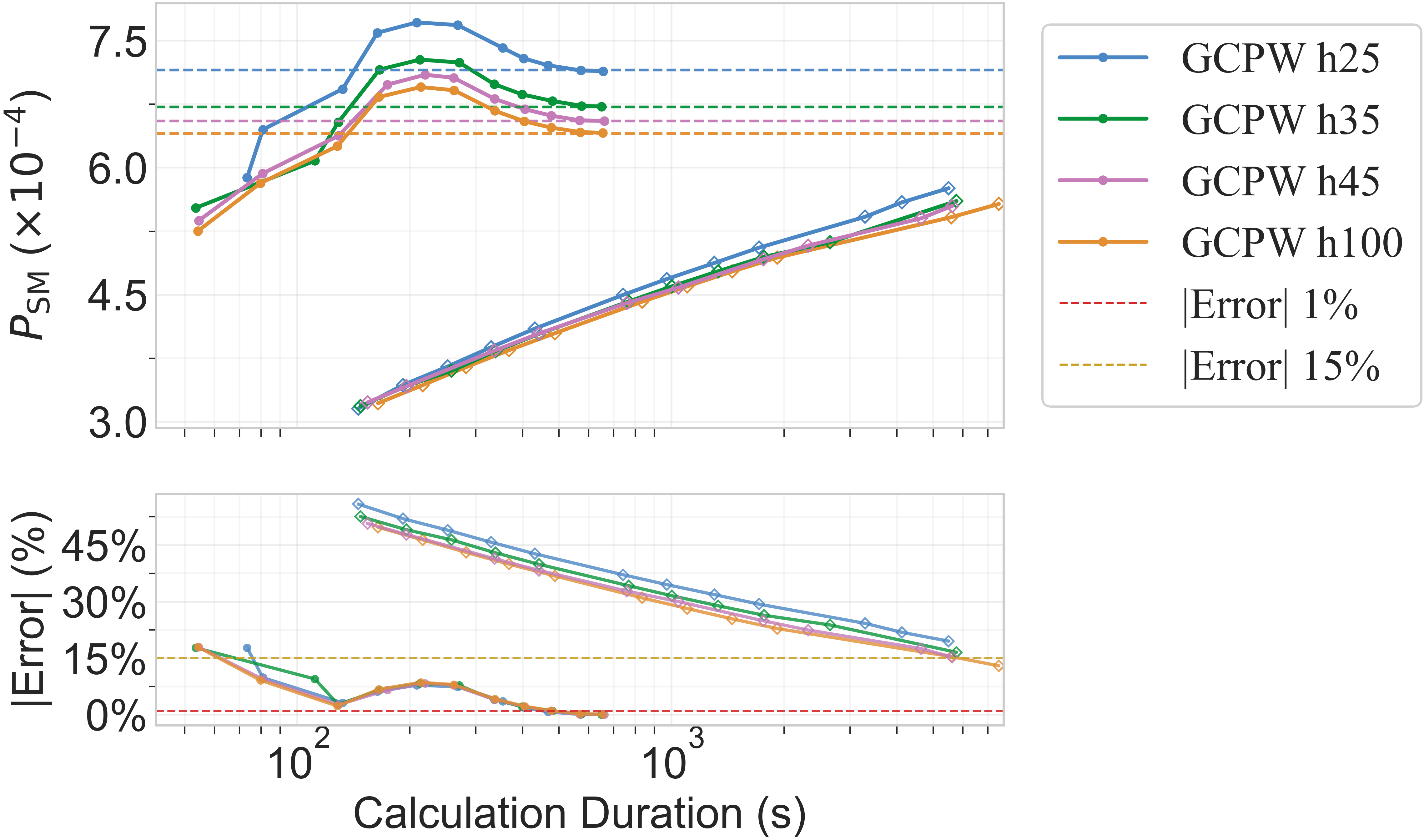} 
    \caption{Comparison of EPR convergence at the SM interface using FEM and SIE. The simulated convergence behaviors of the FEM (diamonds) and SIE (circles) are validated against analytical values (dotted lines) for GCPWs with $a=5\ \mu$m, $b=30\ \mu$m, on a silicon substrate ($\epsilon_{\text{sub}} = 11.9$). A contamination layer ($\epsilon_c = 11.9$) with a thickness of $\delta = 3$ nm is evaluated across four different layer thickness configurations: $h=25\ \mu$m (blue), $h=35\ \mu$m (green), $h=45\ \mu$m (purple), $h=100\ \mu$m (orange). The red dotted line indicates a 1\% relative error boundary, and the yellow dotted line indicates a 15\% relative error boundary. Notably, the FEM fails to converge due to memory exhaustion after approximately 6000~s with an accuracy of $15\%$, whereas the SIE achieves the same accuracy in roughly 100~s,
    demonstrating a computational speedup of two orders of magnitude.}
    \label{fig:combined_gcpw_psm}
\end{figure}

Achieving convergence for the multilayer EPR calculation is computationally demanding, not only due to the multiscale computational challenge, but also due to the scattered field coming from the bottom ground plane.
While the FEM approach necessitates a fine volumetric meshing of tetrahedrons within the intermediate layer, the SIE method requires meshing only along the 2D conductor surfaces. 
As shown in Fig.~\ref{fig:combined_gcpw_psm}, the FEM solution struggles to converge to the analytical result. Even when exhausting the available memory resources, the relative error remains approximately $15\%$.
Conversely, SIE achieves the same accuracy in roughly 50 s, demonstrating a computational speedup of two orders of magnitude.

\subsection{2D Transmon}
\begin{figure}[h]
    \centering 
    \includegraphics[width=\linewidth]{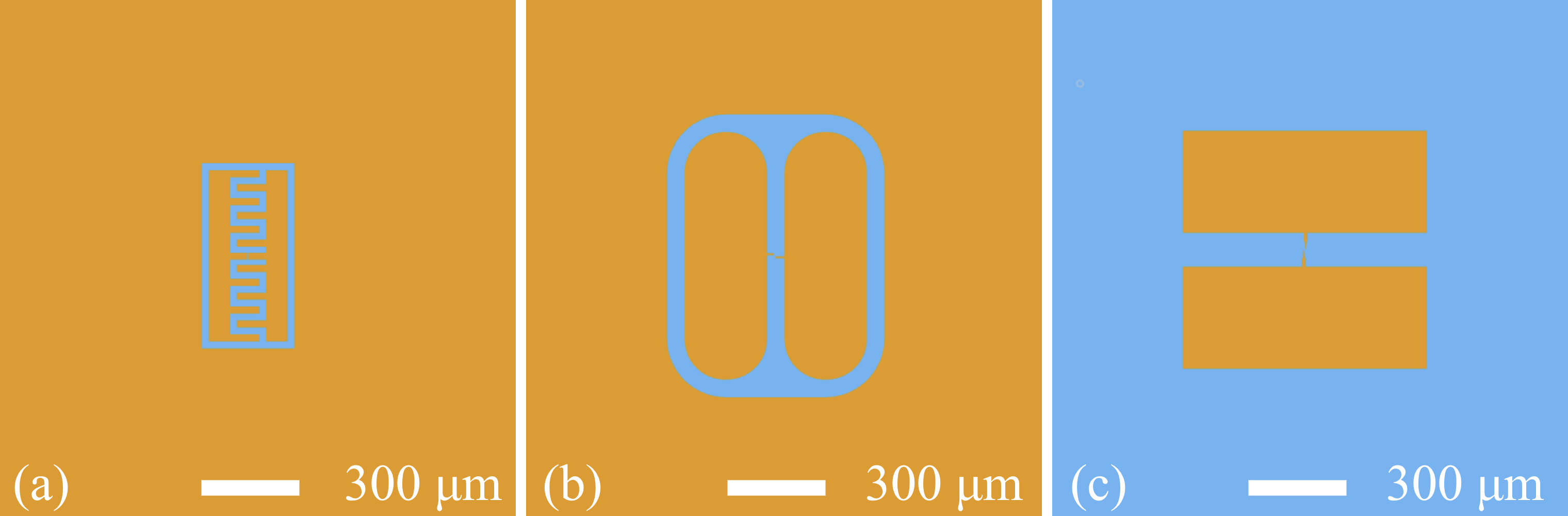} 
    \caption{Layout of typical transmon qubits with different types: (a) 2D interdigital, (b) 2D dumbbell, and (c) 3D dumbbell. The yellow area is the TiN base layer, and the blue area is the sapphire substrate. 
    Layout images are rendered from the simulation GDSII files.
    } 
   \label{figure:transmon_sketch} 
\end{figure}

To demonstrate the proposed method using quantum devices, we investigate three types of transmons, as shown in Fig.~\ref{figure:transmon_sketch}, which are the 2D interdigital transmon, the 2D dumbbell transmon, and the 3D dumbbell transmon.  
The interdigital transmon structure distributes more energy at the substrate-metal (SM) interface~\cite{Gambetta2020}. This type of qubit possesses a larger $P_\text{SM}$, which is tunable over a wide range by adjusting the spacing and number of fingers. 
In contrast, the dumbbell transmon exhibits a significantly lower $P_\text{SM}$. Consequently, the EPR in our study is adjustable within a large range. 
% This tunability facilitates data processing and the fitting of the loss tangent $\tan\delta_c$.

Simulation model parameters are derived from material characterizations using transmission electron microscopy (TEM) in~\cite{Deng2023}. 
TEM images reveal an approximately $1$~nm thick disordered layer between the TiN film and the sapphire substrate. Accordingly, we set the thickness of the SM interface to 
$t_\text{SM} = 1$~nm and its permittivity to $\epsilon_c = 10.15 \epsilon_0$ in the numerical simulations.

We apply both numerical methods to three types of practical transmon qubit designs, denoted as transmon (a), (b), and (c). 
Simulations utilize the commercial FEM simulator ANSYS Maxwell and the proposed SIE method. 
For the FEM simulations, adaptive mesh refinement is enabled with an energy error convergence target of 0.02\%.
For the SIE method, the surface mesh undergoes non-conformal refinement near the edges to capture the charge distribution. 
Table~\ref{tab:transmon_cap_matrix} presents a comparison of the primary components of the capacitance matrices calculated by the two methods.
 
The transmons investigated in this section primarily consist of two plates of a shunting capacitor, denoted as qa and qb. 
The 2D transmon also includes a surrounding ground conductor (denoted as gnd), whereas the 3D transmon does not. 
The two plates are connected by a Josephson junction. We extract the mutual capacitance between the qubit shunting capacitor plates ($C_\text{qa,qb}$) and the capacitance of each plate to ground ($C_\text{qa,gnd}$ and $C_\text{qb,gnd}$) from the capacitance matrix. 
The table indicates a high degree of agreement between the SIE and FEM methods in capacitance calculations, with relative errors remaining within 1\%. This agreement demonstrates that both methods exhibit comparable consistency in 
capacitance solutions.

\begin{table}[h]
\centering
\caption{Comparison of capacitance matrix components and relative errors calculated by the proposed SIE method and the FEM simulator for three transmon designs.}
\label{tab:transmon_cap_matrix}
\begin{tabular}{c|c|ccc}
\hline
Device & Method & $C_\text{qa,qb}$ (fF) & $C_\text{qa,gnd}$ (fF) & $C_\text{qb,gnd}$ (fF) \\
\hline
\multirow{3}{*}{Transmon (a)} & SIE & 54.80 & 79.65 & 72.64 \\
& FEM & 54.90 & 79.68 & 72.67 \\
& Diff. (\%) & 0.18 & 0.04 & 0.04 \\
\hline
\multirow{3}{*}{Transmon (b)} & SIE & 41.90 & 119.82 & 119.87 \\
& FEM & 42.27 & 119.40 & 119.44 \\
& Diff. (\%) & 0.88 & -0.35 & -0.46 \\
\hline
\multirow{3}{*}{Transmon (c)} & SIE & 95.75 & - & - \\
& FEM & 94.95 & - & - \\ 
& Diff. (\%) & -0.84 & - & - \\
\hline
\end{tabular}
\end{table}

Table~\ref{tab:transmon_psm_breakdown} details the comparison of the EPR at the SM interface ($P_\text{SM}$). To analyze the sources of dielectric loss more comprehensively, we decompose the total participation ratio $P_\text{total}$ into contribution components from different conductor surfaces, specifically $P_\text{qa}$, $P_\text{qb}$, and $P_\text{gnd}$.

\begin{table}[h]
\centering\caption{Comparison of the participation ratio ($P_\text{SM}$) and its components calculated by the proposed SIE method and the FEM simulator (unit: $\times 10^{-4}$). $P_\text{total}$ values computed by SIE come from~\cite{Deng2023}.}
\label{tab:transmon_psm_breakdown}
\begin{tabular}{c|c|cccc}
\hline
Device & Method & $P_\text{qa}$ & $P_\text{qb}$ & $P_\text{gnd}$ & $P_\text{total}$~\cite{Deng2023} \\
\hline
\multirow{3}{*}{Transmon (a)} & SIE & 1.04 & 0.98 & 0.10 & 2.12 \\
& FEM & 0.70 & 0.67 & 0.06 & 1.43 \\
& Diff. (\%) & -32.52 & -32.62 & -31.77 & -32.52 \\
\hline
\multirow{3}{*}{Transmon (b)} & SIE & 0.23 & 0.23 & 0.05 & 0.51 \\
& FEM & 0.17 & 0.17 & 0.03 & 0.36 \\
& Diff. (\%) & -26.79 & -27.38 & -39.64 & -28.39 \\
\hline
\multirow{3}{*}{Transmon (c)} & SIE & 0.17 & 0.17 & - & 0.33 \\
& FEM & 0.12 & 0.12 & - & 0.24 \\
& Diff. (\%) & -28.44 & -28.44 & - & -28.44 \\
\hline
\end{tabular}
\end{table}

Several observations arise from the results in Table~\ref{tab:transmon_psm_breakdown}. 
First, the total participation ratio calculated by the SIE method is higher than the FEM results in all test cases, exhibiting a difference of approximately 30\%. 
% Second, further analysis reveals that this discrepancy originates primarily from the contributions of the qa and qb plates. 
Because transmon qubits possess strong fringing field effects, the SIE method captures the electric field singularity at the edges more accurately through its non-conformal mesh and analytical integral solution~\cite{Yla-Oijala2003}. 
In contrast, the FEM method relies on volumetric meshes and tends to underestimate the energy participation at the edges without extreme mesh refinement. 

Furthermore, the contribution of the ground plane ($P_\text{gnd}$) is non-negligible in certain designs such as transmon~(a) and transmon~(b). The SIE method accurately calculates the electric field energy contribution from this region. 

These results indicate that traditional FEM simulations may underestimate EPR when evaluating the qubit electric field. The SIE method proposed in this work provides a more conservative and potentially better converged evaluation approach.

To demonstrate the relationship between EPR and experimental coherence times in real quantum devices,
Ref.~\cite{Deng2023} utilized titanium nitride (TiN) superconducting film transmon qubits on sapphire substrates as an experimental platform. 
% Due to their chemical stability under oxidizing conditions and their capability to form high-quality interfaces on substrates~\cite{Richardson2020}, TiN films on sapphire substrates are considered an ideal material system for realizing superconducting qubits with high coherence times. 
In that study, the authors systematically fabricated multiple sets of transmon qubits on the same batch of TiN base layer,
successfully characterized the quality factors of qubits versus $P_\text{SM}$ and estimated the loss tangent at the SM interface.

\subsection{Optimization of Rectangular Qubit Pattern}
In~\cite{Gambetta2017}, the authors reported engineering the EPR through qubit pattern designs. 
In that work, the EPR was spanned over a wide range to highlight the dielectric losses for material interface studies. 
In this work, we optimize the qubit pattern to improve the EPR, thereby demonstrating
the capability to optimize qubit layouts efficiently.
The capacitive energy $E_c={e^2}/({2C})$ is usually a constraint in the design, 
where $C$ is the capacitance across the Josephson junction in the effective qubit circuit model~\cite{Ding2020}. 

We start with a 3D dumbbell transmon consisting of two identical rectangular pads separated by a fixed distance, 
there is an optimal aspect ratio that minimizes the EPR~\cite{Wang2015}. 
Instead of having charges concentrated in the adjacent regions, 
charges spread out owing to the large partial capacitance contribution from the more distant pad regions. 

\begin{figure}[h]
  \centering 
    \includegraphics[width=0.75\linewidth]{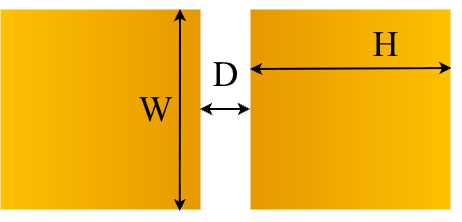} 
    \caption{Schematic diagram of the rectangular qubit structure. The design is parameterized by the inner edge width $W$, the inter-pad gap distance $D$, and the pad extent $H$.
    } 
    \label{figure:rectangular} 
\end{figure} 
\begin{figure}[h]
    \centering 
    \includegraphics[width=\linewidth, trim=0pt 0pt 2pt 0pt, clip]{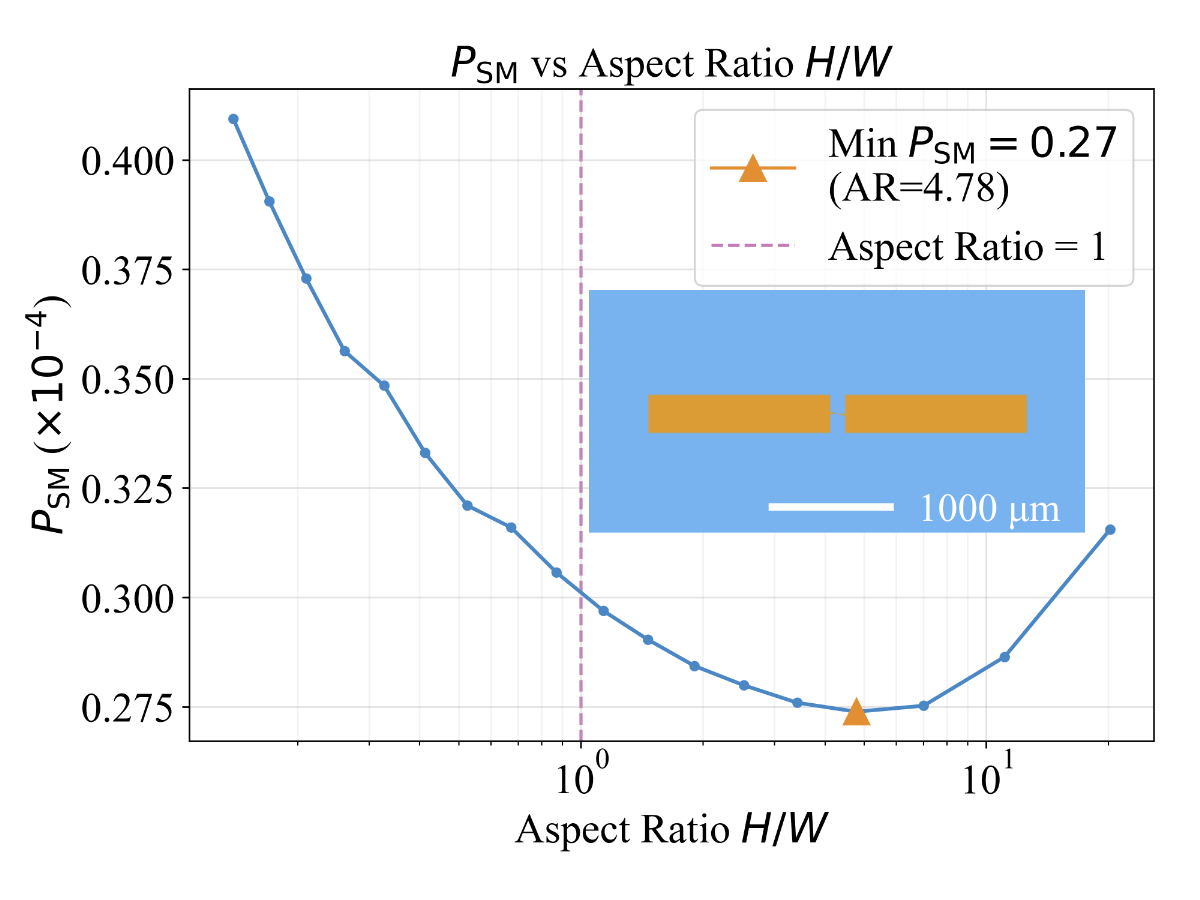} 
    \caption{The EPR at the SM interface of the design with $E_c/h=1.8$ GHz for various values of aspect ratio $H/W$. The solid blue curve illustrates the variation of $P_{\text{SM}}$ highlighting a distinct minimum. A vertical dashed purple line denotes the location of an aspect ratio of 1. The minimum $P_{\text{SM}}$ value of $0.27 \times 10^{-4}$ is reached at $H/W = 4.78$, indicated by the orange triangle. The inset illustrates the layout of the qubit at this optimized aspect ratio.} 
    \label{figure:rectangular_psm} 
\end{figure} 

The numerical simulations are performed on the rectangular qubit with 
fixed $D=8$ $\mu$m and $E_c/h=1.8$ GHz. For a given inner edge width $W$, a specific value of pad extent $H$ can be found that meets the target $E_c$. 
It is again assumed that the dielectric constants of the 
silicon substrate and contamination interface are $\epsilon_{\text{sub}} = \epsilon_c=10.15$ and the interface thickness $\delta$ is 1~nm. 

As shown in Fig.~\ref{figure:rectangular_psm}, the optimal rectangular qubit design achieves a minimum $P_{\text{SM}}$ of $0.27 \times 10^{-4}$ at an aspect ratio of approximately 4.78. The inset illustrates the qubit layout at this optimized geometry. This successful EPR minimization demonstrates SesQ's high efficiency in facilitating rapid iterative layout optimization.
   
\section{Conclusion} ~\label{sec:conclusion}
In this paper, we present SesQ, a highly efficient SIE simulator tailored for evaluating the EPR of superconducting qubits. By applying 2D surface discretization, utilizing the semi-analytical multilayer Green's function, and employing non-conformal boundary mesh refinement, the proposed method overcomes the severe multiscale computational bottlenecks of the conventional FEM simulators. This integral equation approach efficiently captures singular charge density distributions at superconductor edges without an explosive growth in unknowns.

Validations against analytical models demonstrated that SesQ achieves computational accelerations of around two orders of magnitude compared to FEM tools in capacitance extraction. The simulator provides a much more reliable and better-converged evaluation of the EPR, which
is often underestimated by FEM simulations. The simulator's efficiency enables rapid iterative layout optimizations, as demonstrated by the EPR minimization of a rectangular qubit.

Future work will extend the SIE simulator to efficiently simulate magnetic fields. While the magnetostatic field is governed by the vector Poisson equation and its complete solution requires a dyadic Green's function, the $\rho$- and $z$-components of the Green's function in multilayer structures can be reduced to the same form as in the electrostatic case. 
By assigning an excitation current density, the vector magnetic potential can be computed using the numerical techniques developed in this work. 
Consequently, the inductance and magnetic field distribution can be accurately computed, allowing the multiscale bottlenecks traditionally encountered when computing the magnetic field participation ratio to be mitigated using this SIE approach.

Moreover, this work lays the foundation for an end-to-end simulation pipeline. The proposed simulator can be straightforwardly embedded into an automatic differentiation framework. By developing differentiable operators for mesh generation and surface integral equations, gradients can be computed efficiently via backpropagation. 
Moving forward, this will enable the use of gradient-based optimizers to significantly accelerate iterative layout optimization.

\section*{Acknowledgment}
This work is supported by National Key Research and Development Program of China (Grant No. 2025YFE0200900) and Zhongguancun Laboratory. Z. W. and X. W. acknowledge the partial support from the National Key Research and Development Program of China (2021YFA1401902). 
Part of this work was done when Z.W. was visiting Tsinghua University.
We thank Ran Gao, Kun Ding, Shaowei Li, and Chen Zha for providing valuable suggestions. Z. W., F. W., X. T. and H.-H. Z. would like to thank their former colleagues at the Alibaba Quantum Laboratory for helpful discussions.

% ============================================
\appendices

\section{Charge Neutrality}\label{sec:charge_neutrality}
The matrix equation in \eqref{matrix_eqn} is an ill-conditioned system due to the lack of a definition for the total charge. 
% In principle, the total charge at infinity should be zero, 
% which implies that the total charge vanishes in this problem \cite{qian2009fast} \cite{xia2016enhanced}.
This ill-conditioning can be fixed by adding the charge neutrality constraint:
\begin{equation}\label{charge_neutrality}
  \sum_i a_i = 0.
\end{equation}

One way to apply the charge neutrality constraint in the matrix system is to 
first subtract the $N$-th equation in \eqref{matrix_eqn} 
from each of the other $N-1$ equations. Then replace the $N$-th equation with the constraint from~\eqref{charge_neutrality}. 
This yields:
% Denoting the elements of the new matrix $\mat{G}^\prime$ and the new potential vector $\vg{\phi}$ become:
\begin{equation}
\left\{
\begin{aligned}
    \left[\vg{\phi}^\prime\right]_i &= \left[\vg{\phi}\right]_i - \left[\vg{\phi}\right]_N, \\
    \left[\mat{G}^\prime\right]_{ij} &= \left[\mat{G}\right]_{ij} - \left[\mat{G}\right]_{Nj},
\end{aligned}
\right. \quad \text{for } i < N,
\end{equation}
and
% which designates the $N$-th element as the reference.
% Then replace the $N$-th equation with the constraint from~\eqref{charge_neutrality}, yielding
\begin{equation}
\left\{
\begin{aligned}
    \left[\vg{\phi}^\prime\right]_N &= 0, \\
    \left[\mat{G}^\prime\right]_{Nj} &= 1,
\end{aligned}
\right. \quad \text{for all } j.
\end{equation}
Finally, a modified well-conditioned matrix equation can be formulated as:
\begin{equation}\label{matrix_eqn2}
  \mat{G}^\prime \cdot \v{q} = \vg{\phi}^\prime.
  % \vg{\phi} 
\end{equation}

\section{Preparing the Green's function database} \label{sec:prepare_database}
Recalling the form of the integral equation~\eqref{integral_form}, the integration is performed over the superconductor surfaces $\mathcal{S}$ for all source points. Consequently, many instances of the Green's function are evaluated repeatedly for the same distances.
Due to the high computational cost of evaluating the multilayer Green's function using numerical integration methods, it is highly beneficial to cache the Green's function points in a database.
During the construction of the integral equation matrix, if $G(\rho, z;z^\prime)$ falls within the cached distance range of the database, we can rapidly evaluate it using a fast interpolation method rather than executing the full numerical procedure again.

\section{Derivation of energy participation ratio approximation formula} \label{sec:psm_intermediate_layer_derivation}
Based on the conformal mapping method, the electric field of grounded CPW is analytically solvable, which takes the forms~\eqref{gcpw_analytical_electric_field_out} and~\eqref{gcpw_analytical_electric_field_in}.
We derive an approximate closed-form EPR solution using the electric field within the intermediate layer.
To simplify the derivation, dimensionless variables $X = \pi x/(2h)$, $A = \pi a/(2h)$, and $B = \pi b/(2h)$ are introduced. The electric field distribution $E_\text{in}(\zeta)$ along the interface $z \to 0$, can be written as:
\begin{equation}
\begin{aligned}
    A_0&=\frac{\phi_0 \pi \cosh A \cdot \sinh B}{2h K(k_1')},\\
    E_\text{in}(\zeta)&\approx A_0\frac{1}{\sqrt{[\cosh^2 X-\cosh^2 A] [\cosh^2 X-\cosh^2 B]}},
\end{aligned}
\end{equation}
where $k_1$ is the geometric argument of the intermediate region, and $K(k_1')$ is the complete elliptic integral of the first kind. 

At a small height $z$ close to the interface (satisfying $z \ll a, b, h$), the square of the electric field modulus $\theta(x,z)$, can be approximated as:
\begin{equation}
\theta(x,z) \approx \frac{A_0^2}{(\cosh^2 B - \cosh^2 X)(\cosh^2 A - \cosh^2 X)}.
\end{equation}
Since the electric field exhibits singularities at the conductor edges $x \to a$ and $x \to b$, direct integration is rather difficult. Therefore, a singularity extraction method is employed. Using a partial fraction expansion:
\begin{equation}
\theta(x,z) \approx C_\text{pre} \left[ \frac{1}{\cosh^2A - \cosh^2X} - \frac{1}{\cosh^2B - \cosh^2X} \right],
\end{equation}
where the prefactor is
\begin{equation}
C_\text{pre} = \frac{A_0^2}{\cosh^2 B - \cosh^2 A}.
\end{equation}

Near the signal line edge ($x \to a$), the primary integral contribution comes from the singular term. 
The singular part $\theta_a^s(x,z)$ can be expressed as:
\begin{equation}
\theta_a^s(x,z) \approx C_\text{pre} \frac{1}{\cosh^2A - \cosh^2X}.
\end{equation}
Integrating the singular term over the non-singular region $[0, a-\delta]$ gives, 
\begin{equation}
\int_0^{a-\delta} \theta_a^s(x,z) dx = \frac{2h C_\text{pre}}{\pi \sinh(2A)}\ln \left( \frac{2h \sinh(2A)}{\pi \delta} \right),
\end{equation}
the $z$-regularized integral of the singular portion over region $[a-\delta, a]$ evaluates to 
\begin{equation}
\int_{a-\delta}^a \theta_a^s(x,z) dx = \frac{2h C_\text{pre}}{\pi \sinh(2A)}\ln \left( \frac{2\delta}{z} \right).
\end{equation}
Hence, summing two regions yields the cancellation of the arbitrary cutoff $\delta$. 
Combining it with the integration result of the non-singular term, 
% $\ln \left[ {\sinh(B-A)}/{\sinh(B+A)} \right] \cdot {2h}/\left[{\pi \sinh(2B)}\right]$, 
yields the total contribution at the signal line edge:
\begin{equation}
\begin{aligned}
    \int_0^a \theta_a(x,z) dx \approx \frac{2h C_\text{pre}}{\pi} &\left[ \frac{1}{\sinh(2A)} \ln\left( \frac{4h \sinh(2A)}{\pi z} \right) \right. \\
    &\left. +\frac{1}{ \sinh(2B)} \ln \frac{\sinh(B-A)}{\sinh(B+A)}\right].
\end{aligned}
\end{equation}

Similarly, near the ground plane edge ($x \to b$), integrating singular term $\theta_b^s(x,z)$ over the interval $[b, +\infty)$ and combining the contributions of the singular and non-singular terms yields:
\begin{equation}
\begin{aligned}
\int_b^\infty \theta_b(x,z) &dx \approx \\
&\frac{2h C_\text{pre}}{\pi} \cdot \left[ \frac{1}{\sinh(2A)}\left( 2A + \ln \frac{\sinh(B-A)}{\sinh(B+A)} \right) \right.\\
&\quad \left. + \frac{1}{\sinh(2B)}\ln\left( \frac{4h \exp(-2B) \sinh(2B)}{\pi z} \right)\right].
\end{aligned}
\end{equation}
Combining the integration results from both parts provides the total surface energy density function $\Theta(z)$ as it varies with depth $z$,
\begin{equation}
    \Theta(z) = \left[ \int_{0}^{a} \theta_a(x, z) dx + \int_{b}^{\infty} \theta_b(x, z) dx \right],
\end{equation}
note that the factor of 2 from the left/right geometric symmetry of the GCPW cancels out with the $1/2$ from the electric energy density formula.
To calculate the energy participation ratio, the total energy $U_\text{tot} = C_\text{gcpw}\phi_0^2/2$ is introduced.

Substituting $C_\text{pre}$ and simplifying using the geometric relation:
\begin{equation}
    \frac{\sinh(B-A)}{\sinh(B+A)} = \frac{1-k_1}{1+k_1},
\end{equation}
the participation ratio density satisfies:
\begin{equation}
\begin{aligned}
    r(z) =& \frac{\epsilon_0 \epsilon_\text{in}^2 \Theta(z)}{\epsilon_c U_\text{tot}}\\
    =&\frac{\epsilon_0\epsilon_\text{in}^2}{\epsilon_c C_\text{gcpw}} \cdot \frac{\pi}{h [k_1' K(k_1')]^2} \left( \frac{\Lambda_a(z)}{\sinh(\frac{\pi a}{h})} + \frac{\Lambda_b(z)}{\sinh(\frac{\pi b}{h})} \right),
\end{aligned}
\end{equation}
where $\Lambda_a(z)$ and $\Lambda_b(z)$ satisfy:
\begin{equation}
\begin{aligned}
    \Lambda_a &= \ln\left( \frac{4h \sinh(\frac{\pi a}{h})}{\pi z} \right) + \ln\left( \frac{1-k_1}{1+k_1} \right) + \frac{\pi a}{h},\\
    \Lambda_b &= \ln\left( \frac{4h \exp\left( -\frac{\pi b}{h} \right) \sinh(\frac{\pi b}{h})}{\pi z} \right) + \ln\left( \frac{1-k_1}{1+k_1} \right).
\end{aligned}
\end{equation}

Finally, integrating over a thin surface layer of thickness $\delta$, 
\begin{equation}
    P_\text{SM}(\delta) \approx \int_0^{\delta}r(z)dz, 
\end{equation}
yields the analytical expression for the energy participation ratio of the surface interface.

\ifCLASSOPTIONcaptionsoff
  \newpage
\fi



%\begin{thebibliography}{1}


\bibliographystyle{IEEEtran}
\bibliography{IEEEabrv,Bibliography}

\begin{thebibliography}{10}
\providecommand{\url}[1]{#1}
\csname url@rmstyle\endcsname
\providecommand{\newblock}{\relax}
\providecommand{\bibinfo}[2]{#2}
\providecommand\BIBentrySTDinterwordspacing{\spaceskip=0pt\relax}
\providecommand\BIBentryALTinterwordstretchfactor{4}
\providecommand\BIBentryALTinterwordspacing{\spaceskip=\fontdimen2\font plus
\BIBentryALTinterwordstretchfactor\fontdimen3\font minus \fontdimen4\font\relax}
\providecommand\BIBforeignlanguage[2]{{%
\expandafter\ifx\csname l@#1\endcsname\relax
\typeout{** WARNING: IEEEtran.bst: No hyphenation pattern has been}%
\typeout{** loaded for the language `#1'. Using the pattern for}%
\typeout{** the default language instead.}%
\else
\language=\csname l@#1\endcsname
\fi
#2}}

\bibitem{Koch2007}
\BIBentryALTinterwordspacing
J.~Koch, T.~M. Yu, J.~Gambetta, A.~A. Houck, D.~I. Schuster, J.~Majer, A.~Blais, M.~H. Devoret, S.~M. Girvin, and R.~J. Schoelkopf, ``Charge-insensitive qubit design derived from the cooper pair box,'' \emph{Phys. Rev. A}, vol.~76, p. 042319, Oct 2007. [Online]. Available: \url{https://doi.org/10.1103/PhysRevA.76.042319}
\BIBentrySTDinterwordspacing

\bibitem{Martinis2005}
\BIBentryALTinterwordspacing
J.~M. Martinis, K.~B. Cooper, R.~McDermott, M.~Steffen, M.~Ansmann, K.~D. Osborn, K.~Cicak, S.~Oh, D.~P. Pappas, R.~W. Simmonds, and C.~C. Yu, ``{Decoherence in Josephson Qubits from Dielectric Loss},'' \emph{Phys. Rev. Lett.}, vol.~95, no.~21, p. 210503, nov 2005. [Online]. Available: \url{https://doi.org/10.1103/PhysRevLett.95.210503}
\BIBentrySTDinterwordspacing

\bibitem{Wang2015}
\BIBentryALTinterwordspacing
C.~Wang, C.~Axline, Y.~Y. Gao, T.~Brecht, Y.~Chu, L.~Frunzio, M.~H. Devoret, and R.~J. Schoelkopf, ``{Surface participation and dielectric loss in superconducting qubits},'' \emph{Appl. Phys. Lett.}, vol. 107, no.~16, p. 162601, 2015. [Online]. Available: \url{https://doi.org/10.1063/1.4934486}
\BIBentrySTDinterwordspacing

\bibitem{Muller2019}
\BIBentryALTinterwordspacing
C.~Müller, J.~H. Cole, and J.~Lisenfeld, ``Towards understanding two-level-systems in amorphous solids: insights from quantum circuits,'' \emph{Rep. Prog. Phys.}, vol.~82, no.~12, p. 124501, Oct. 2019. [Online]. Available: \url{https://doi.org/10.1088/1361-6633/ab3a7e}
\BIBentrySTDinterwordspacing

\bibitem{Gambetta2017}
\BIBentryALTinterwordspacing
J.~M. Gambetta, C.~E. Murray, Y.~K. Fung, D.~T. Mcclure, O.~Dial, W.~Shanks, J.~W. Sleight, and M.~Steffen, ``{Investigating surface loss effects in superconducting transmon qubits},'' \emph{IEEE Trans. Appl. Supercond.}, vol.~27, no.~1, pp. 1--5, 2017. [Online]. Available: \url{https://doi.org/10.1109/TASC.2016.2629670}
\BIBentrySTDinterwordspacing

\bibitem{Minev2021}
\BIBentryALTinterwordspacing
Z.~K. Minev, Z.~Leghtas, S.~O. Mundhada, L.~Christakis, I.~M. Pop, and M.~H. Devoret, ``Energy-participation quantization of josephson circuits,'' \emph{npj Quantum Inf.}, vol.~7, no.~1, p. 131, 2021. [Online]. Available: \url{https://doi.org/10.1038/s41534-021-00461-8}
\BIBentrySTDinterwordspacing

\bibitem{Lisenfeld2019}
\BIBentryALTinterwordspacing
J.~Lisenfeld, A.~Bilmes, A.~Megrant, R.~Barends, J.~Kelly, P.~Klimov, G.~Weiss, J.~M. Martinis, and A.~V. Ustinov, ``Electric field spectroscopy of material defects in transmon qubits,'' \emph{npj Quantum Inf.}, vol.~5, no.~1, p. 105, Nov. 2019. [Online]. Available: \url{https://doi.org/10.1038/s41534-019-0224-1}
\BIBentrySTDinterwordspacing

\bibitem{Eun2023}
\BIBentryALTinterwordspacing
S.~Eun, S.~H. Park, K.~Seo, K.~Choi, and S.~Hahn, ``Shape optimization of superconducting transmon qubits for low surface dielectric loss,'' \emph{Journal of Physics D: Applied Physics}, vol.~56, no.~50, p. 505306, sep 2023. [Online]. Available: \url{https://doi.org/10.1088/1361-6463/acf7cf}
\BIBentrySTDinterwordspacing

\bibitem{Gao2008}
\BIBentryALTinterwordspacing
J.~Gao, M.~Daal, A.~Vayonakis, S.~Kumar, J.~Zmuidzinas, B.~Sadoulet, B.~A. Mazin, P.~K. Day, and H.~G. Leduc, ``Experimental evidence for a surface distribution of two-level systems in superconducting lithographed microwave resonators,'' \emph{Appl. Phys. Lett.}, vol.~92, no.~15, p. 152505, Apr. 2008. [Online]. Available: \url{https://doi.org/10.1063/1.2906373}
\BIBentrySTDinterwordspacing

\bibitem{Dial2016}
\BIBentryALTinterwordspacing
O.~Dial, D.~T. McClure, S.~Poletto, G.~A. Keefe, M.~B. Rothwell, J.~M. Gambetta, D.~W. Abraham, J.~M. Chow, and M.~Steffen, ``Bulk and surface loss in superconducting transmon qubits,'' \emph{Supercond. Sci. Technol.}, vol.~29, no.~4, p. 044001, Apr. 2016. [Online]. Available: \url{https://doi.org/10.1088/0953-2048/29/4/044001}
\BIBentrySTDinterwordspacing

\bibitem{Ganjam2024}
\BIBentryALTinterwordspacing
S.~Ganjam, Y.~Wang, Y.~Lu, A.~Banerjee, C.~U. Lei, L.~Krayzman, K.~Kisslinger, C.~Zhou, R.~Li, Y.~Jia, M.~Liu, L.~Frunzio, and R.~J. Schoelkopf, ``Surpassing millisecond coherence in on chip superconducting quantum memories by optimizing materials and circuit design,'' \emph{Nat. Commun.}, vol.~15, no.~1, p. 3687, May 2024. [Online]. Available: \url{https://doi.org/10.1038/s41467-024-47857-6}
\BIBentrySTDinterwordspacing

\bibitem{Wenner2011}
\BIBentryALTinterwordspacing
J.~Wenner, R.~Barends, R.~C. Bialczak, Y.~Chen, J.~Kelly, E.~Lucero, M.~Mariantoni, A.~Megrant, P.~J.~J. O'Malley, D.~Sank, A.~Vainsencher, H.~Wang, T.~C. White, Y.~Yin, J.~Zhao, A.~N. Cleland, and J.~M. Martinis, ``Surface loss simulations of superconducting coplanar waveguide resonators,'' \emph{Appl. Phys. Lett.}, vol.~99, no.~11, p. 113513, sep 2011. [Online]. Available: \url{https://doi.org/10.1063/1.3637047}
\BIBentrySTDinterwordspacing

\bibitem{Martinis2022}
\BIBentryALTinterwordspacing
J.~M. Martinis, ``Surface loss calculations and design of a superconducting transmon qubit with tapered wiring,'' \emph{npj Quantum Inf.}, vol.~8, no.~1, p.~26, Mar. 2022. [Online]. Available: \url{https://doi.org/10.1038/s41534-022-00530-6}
\BIBentrySTDinterwordspacing

\bibitem{Elkin2025}
\BIBentryALTinterwordspacing
S.~T. Elkin, G.~Khan, E.~Forati, B.~W. Langley, D.~Timucin, R.~Molavi, S.~Sussman, and T.~E. Roth, ``Opportunities and challenges of computational electromagnetics methods for superconducting circuit quantum device modeling: A practical review,'' 2025. [Online]. Available: \url{https://arxiv.org/abs/2511.20774}
\BIBentrySTDinterwordspacing

\bibitem{Wu2025}
\BIBentryALTinterwordspacing
F.~Wu, J.~Guo, T.~Xia, L.~Kong, F.~Zhang, Z.~Wang, A.~Dai, Z.~Wang, Z.~Yang, H.~Deng, K.~Zhang, Z.~Ji, Y.~Feng, H.-H. Zhao, and J.~Chen, ``Quantum design automation: Foundations, challenges, and the road ahead,'' 2025. [Online]. Available: \url{https://arxiv.org/abs/2511.10479}
\BIBentrySTDinterwordspacing

\bibitem{Murray2018}
\BIBentryALTinterwordspacing
C.~E. Murray, J.~M. Gambetta, D.~T. McClure, and M.~Steffen, ``{Analytical determination of participation in superconducting coplanar architectures},'' \emph{IEEE Trans. Microw. Theory Techn.}, vol.~66, no.~8, pp. 3724--3733, 2018. [Online]. Available: \url{https://doi.org/10.1109/TMTT.2018.2841829}
\BIBentrySTDinterwordspacing

\bibitem{Murray2020}
\BIBentryALTinterwordspacing
C.~E. Murray, ``Analytical modeling of participation reduction in superconducting coplanar resonator and qubit designs through substrate trenching,'' \emph{IEEE Trans. Microw. Theory Techn.}, vol.~68, no.~8, pp. 3263--3270, Aug. 2020. [Online]. Available: \url{https://doi.org/10.1109/TMTT.2020.2995894}
\BIBentrySTDinterwordspacing

\bibitem{Park2023}
\BIBentryALTinterwordspacing
S.~H. Park, D.~Baek, I.~Park, and S.~Hahn, ``Design of scalable superconducting quantum circuits using flip-chip assembly,'' \emph{IEEE Trans. Appl. Supercond.}, vol.~33, no.~5, pp. 1--6, 2023. [Online]. Available: \url{https://doi.org/10.1109/TASC.2023.3244142}
\BIBentrySTDinterwordspacing

\bibitem{Smirnov2024}
\BIBentryALTinterwordspacing
N.~S. Smirnov, E.~A. Krivko, A.~A. Solovyova, A.~I. Ivanov, and I.~A. Rodionov, ``Wiring surface loss of a superconducting transmon qubit,'' \emph{Sci. Rep.}, vol.~14, no.~1, p. 7326, Mar. 2024. [Online]. Available: \url{https://doi.org/10.1038/s41598-024-57248-y}
\BIBentrySTDinterwordspacing

\bibitem{Harrington1993}
R.~F. Harrington, \emph{Field Computation by Moment Methods}.\hskip 1em plus 0.5em minus 0.4em\relax Piscataway, NJ: IEEE Press, 1993.

\bibitem{Gibson2014}
\BIBentryALTinterwordspacing
W.~C. Gibson, \emph{The Method of Moments in Electromagnetics}.\hskip 1em plus 0.5em minus 0.4em\relax CRC Press, 2021. [Online]. Available: \url{https://doi.org/10.1201/9780429355509}
\BIBentrySTDinterwordspacing

\bibitem{Chew1995}
W.~C. Chew, \emph{Waves and fields in inhomogeneous media}.\hskip 1em plus 0.5em minus 0.4em\relax IEEE Press, 1995.

\bibitem{Li2021}
\BIBentryALTinterwordspacing
X.~Li, I.~Jeffrey, M.~Al-Qedra, and V.~I. Okhmatovski, ``Error-{Controlled} {Static} {Layered}-{Medium} {Green}'s {Function} {Computation} via \textit{hp} -{Adaptive} {Spectral} {Differential} {Equation} {Approximation} {Method},'' \emph{IEEE Trans. Compon. Packag. Manuf. Technol.}, vol.~11, no.~9, pp. 1329--1342, Sept. 2021. [Online]. Available: \url{https://doi.org/10.1109/TCPMT.2021.3097944}
\BIBentrySTDinterwordspacing

\bibitem{Michalski1997}
\BIBentryALTinterwordspacing
K.~A. Michalski and J.~R. Mosig, ``Multilayered media green's functions in integral equation formulations,'' \emph{IEEE Trans. Antennas Propag.}, vol.~45, no.~3, pp. 508--519, 1997. [Online]. Available: \url{https://doi.org/10.1109/8.558666}
\BIBentrySTDinterwordspacing

\bibitem{Li2022}
\BIBentryALTinterwordspacing
X.~Li, S.~Zheng, I.~Jeffrey, and V.~I. Okhmatovski, ``Closed-{Form} {Evaluation} of {Mixed} {Potential} {Shielded} {Layered} {Media} {Green}'s {Functions} {With} {Spectral} {Differential} {Equation} {Approximation} {Method},'' \emph{IEEE Trans. Microw. Theory Techn.}, vol.~70, no.~5, pp. 2553--2565, May 2022. [Online]. Available: \url{https://doi.org/10.1109/TMTT.2022.3156917}
\BIBentrySTDinterwordspacing

\bibitem{Ogata2005}
\BIBentryALTinterwordspacing
H.~Ogata, ``A numerical integration formula based on the bessel functions,'' \emph{Publications of the Research Institute for Mathematical Sciences}, vol.~41, no.~4, pp. 949--970, 2005. [Online]. Available: \url{https://doi.org/10.2977/PRIMS/1145474602}
\BIBentrySTDinterwordspacing

\bibitem{Yla-Oijala2003}
\BIBentryALTinterwordspacing
P.~Yla-Oijala and M.~Taskinen, ``{Calculation of CFIE impedance matrix elements with RWG and n x RWG functions},'' \emph{IEEE Trans. Antennas Propag.}, vol.~51, no.~8, pp. 1837--1846, aug 2003. [Online]. Available: \url{https://doi.org/10.1109/TAP.2003.814745}
\BIBentrySTDinterwordspacing

\bibitem{qian2009fast}
\BIBentryALTinterwordspacing
Z.-G. Qian and W.~C. Chew, ``Fast full-wave surface integral equation solver for multiscale structure modeling,'' \emph{IEEE Transactions on Antennas and Propagation}, vol.~57, no.~11, pp. 3594--3601, Nov. 2009. [Online]. Available: \url{https://doi.org/10.1109/TAP.2009.2023629}
\BIBentrySTDinterwordspacing

\bibitem{xia2016enhanced}
\BIBentryALTinterwordspacing
T.~Xia, H.~Gan, M.~Wei, W.~C. Chew, H.~Braunisch, Z.~Qian, K.~Aygun, and A.~Aydiner, ``An enhanced augmented electric-field integral equation formulation for dielectric objects,'' \emph{IEEE Transactions on Antennas and Propagation}, vol.~64, no.~6, pp. 2339--2347, June 2016. [Online]. Available: \url{https://doi.org/10.1109/TAP.2016.2537389}
\BIBentrySTDinterwordspacing

\bibitem{Meixner1972}
\BIBentryALTinterwordspacing
J.~Meixner, ``The behavior of electromagnetic fields at edges,'' \emph{IEEE Trans. Antennas Propag.}, vol.~20, no.~4, pp. 442--446, 1972. [Online]. Available: \url{https://doi.org/10.1109/TAP.1972.1140243}
\BIBentrySTDinterwordspacing

\bibitem{Abramowitz1964}
M.~Abramowitz and I.~A. Stegun, \emph{Handbook of Mathematical Functions with Formulas, Graphs, and Mathematical Tables}.\hskip 1em plus 0.5em minus 0.4em\relax US Government printing office, 1964.

\bibitem{AnsysMaxwell}
\BIBentryALTinterwordspacing
{Ansys, Inc.}, ``Ansys maxwell: Low frequency electromagnetic field simulation,'' 2023. [Online]. Available: \url{https://www.ansys.com/products/electronics/ansys-maxwell}
\BIBentrySTDinterwordspacing

\bibitem{Garg2013}
I.~Bahl, M.~Bozzi, and R.~Garg, \emph{Microstrip Lines and Slotlines, Third Edition}.\hskip 1em plus 0.5em minus 0.4em\relax Artech, 2013.

\bibitem{Sandberg2012}
\BIBentryALTinterwordspacing
M.~Sandberg, M.~R. Vissers, J.~S. Kline, M.~Weides, J.~Gao, D.~S. Wisbey, and D.~P. Pappas, ``Etch induced microwave losses in titanium nitride superconducting resonators,'' \emph{Appl. Phys. Lett.}, vol. 100, no.~26, p. 262605, 06 2012. [Online]. Available: \url{https://doi.org/10.1063/1.4729623}
\BIBentrySTDinterwordspacing

\bibitem{Richmond1991}
\BIBentryALTinterwordspacing
J.~Richmond, ``On the variational aspects of the moment method (electromagnetics),'' \emph{IEEE Trans. Antennas Propag.}, vol.~39, no.~4, pp. 473--479, 1991. [Online]. Available: \url{https://doi.org/10.1109/8.81459}
\BIBentrySTDinterwordspacing

\bibitem{Ghione1983}
\BIBentryALTinterwordspacing
G.~Ghione and C.~U. Naldi, ``Coplanar waveguides for mmic applications: effect of upper shielding, conductor backing, finite-extent ground planes, and line-to-line coupling,'' \emph{IEEE Trans. Microw. Theory Techn.}, vol.~35, no.~3, pp. 260--267, 1987. [Online]. Available: \url{https://doi.org/10.1109/TMTT.1987.1133637}
\BIBentrySTDinterwordspacing

\bibitem{Hanna1985}
\BIBentryALTinterwordspacing
V.~F. Hanna, ``Parameters of coplanar diretional couplers with lower ground plane,'' in \emph{1985 15th European Microwave Conference}, 1985, pp. 820--825. [Online]. Available: \url{https://doi.org/10.1109/EUMA.1985.333579}
\BIBentrySTDinterwordspacing

\bibitem{Li2018}
\BIBentryALTinterwordspacing
Q.~Li and Y.~Zhang, ``Analytical formulas for conductor-backed connected-ground coplanar waveguide,'' in \emph{2018 IEEE International Conference on Computational Electromagnetics (ICCEM)}, 2018, pp. 1--3. [Online]. Available: \url{https://doi.org/10.1109/COMPEM.2018.8496585}
\BIBentrySTDinterwordspacing

\bibitem{Gillick1993}
\BIBentryALTinterwordspacing
M.~Gillick, I.~Robertson, and J.~Joshi, ``Direct analytical solution for the electric field distribution at the conductor surfaces of coplanar waveguides,'' \emph{IEEE Trans. Microw. Theory Techn.}, vol.~41, no.~1, pp. 129--135, 1993. [Online]. Available: \url{https://doi.org/10.1109/22.210239}
\BIBentrySTDinterwordspacing

\bibitem{Piessens1983}
\BIBentryALTinterwordspacing
R.~Piessens, E.~de~Doncker-Kapenga, C.~W. {\"U}berhuber, and D.~K. Kahaner, \emph{QUADPACK: A subroutine package for automatic integration}.\hskip 1em plus 0.5em minus 0.4em\relax Springer Science \& Business Media, 1983. [Online]. Available: \url{https://doi.org/10.1007/978-3-642-61786-7}
\BIBentrySTDinterwordspacing

\bibitem{Gambetta2020}
\BIBentryALTinterwordspacing
J.~M. Gambetta, J.~M. Chow, and M.~Steffen, ``Building logical qubits in a superconducting quantum computing system,'' \emph{npj Quantum Inf.}, vol.~3, no.~1, p.~2, 2017. [Online]. Available: \url{https://doi.org/10.1038/s41534-016-0004-0}
\BIBentrySTDinterwordspacing

\bibitem{Deng2023}
\BIBentryALTinterwordspacing
H.~Deng, Z.~Song, R.~Gao, T.~Xia, F.~Bao, X.~Jiang, H.-S. Ku, Z.~Li, X.~Ma, J.~Qin, H.~Sun, C.~Tang, T.~Wang, F.~Wu, W.~Yu, G.~Zhang, X.~Zhang, J.~Zhou, X.~Zhu, Y.~Shi, H.-H. Zhao, and C.~Deng, ``Titanium nitride film on sapphire substrate with low dielectric loss for superconducting qubits,'' \emph{Phys. Rev. Applied}, vol.~19, p. 024013, Feb 2023. [Online]. Available: \url{https://doi.org/10.1103/PhysRevApplied.19.024013}
\BIBentrySTDinterwordspacing

\bibitem{Ding2020}
\BIBentryALTinterwordspacing
D.~Ding, H.-S. Ku, Y.~Shi, and H.-H. Zhao, ``Free-mode removal and mode decoupling for simulating general superconducting quantum circuits,'' \emph{Phys. Rev. B}, vol. 103, p. 174501, May 2021. [Online]. Available: \url{https://doi.org/10.1103/PhysRevB.103.174501}
\BIBentrySTDinterwordspacing

\end{thebibliography}
% \end{thebibliography}
%\begin{biography}[{\includegraphics[width=1in,height=1.25in,clip,keepaspectratio]{mshell}}]{Michael Shell}

% ==== SWITCH OFF the BIO for submission
% ==== SWITCH OFF the BIO for submission
% \begin{IEEEbiography}[{\includegraphics[width=1in,height=1.25in,clip,keepaspectratio]{photo/mike.png}}]{Michael Roberg}
% (S'09) received the B.S.E.E degree from Bucknell University, Lewisburg, PA, in 2003, the M.S.E.E. degree from the University of Pennsylvania, Philadelphia, in 2006, and the Ph.D. degree from the University of Colorado at Boulder in 2012. From 2003 to 2009, he was an Engineer with Lockheed Martin–MS2, Moorestown, NJ, where he was involved with advanced phased-array radar systems. His current research interests include high efficiency microwave PA theory and design, microwave power rectifiers, MMIC design, and high-efficiency radar and communication system transmitters. He is currently employed by TriQuint Semiconductor - Defense Products and Foundry Services in Richardson, TX working on wideband high efficiency GaN MMIC PA design.
% \end{IEEEbiography}

\begin{IEEEbiography}[{\includegraphics[width=1in,height=1.25in,clip,keepaspectratio]{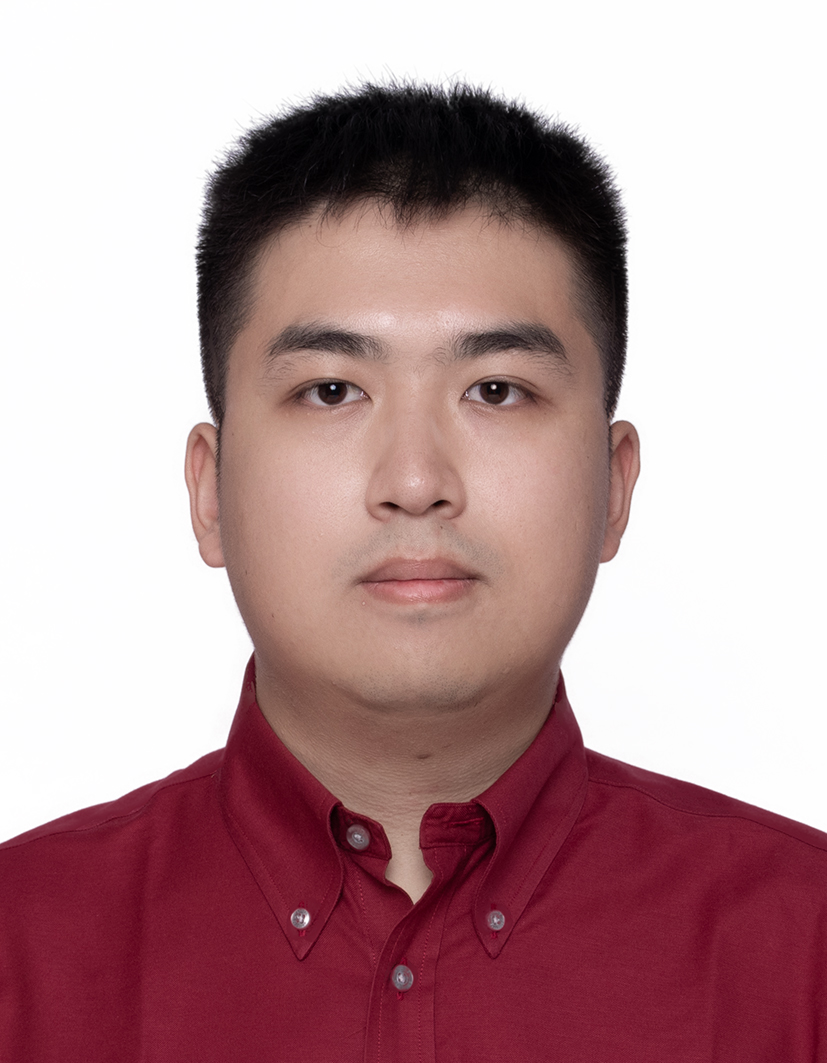}}]{Ziang Wang}
is currently a Ph.D. student in the Zhejiang Institute of Modern Physics, Zhejiang University, Hangzhou. His research focuses on the simulation and optimization of superconducting quantum processors. Before that, he received his bachelor's degree from School of Physics, University of Electronic Science and Technology of China in 2021.
\end{IEEEbiography}

\begin{IEEEbiography}
[{\includegraphics[width=1in,height=1.25in,clip,keepaspectratio]{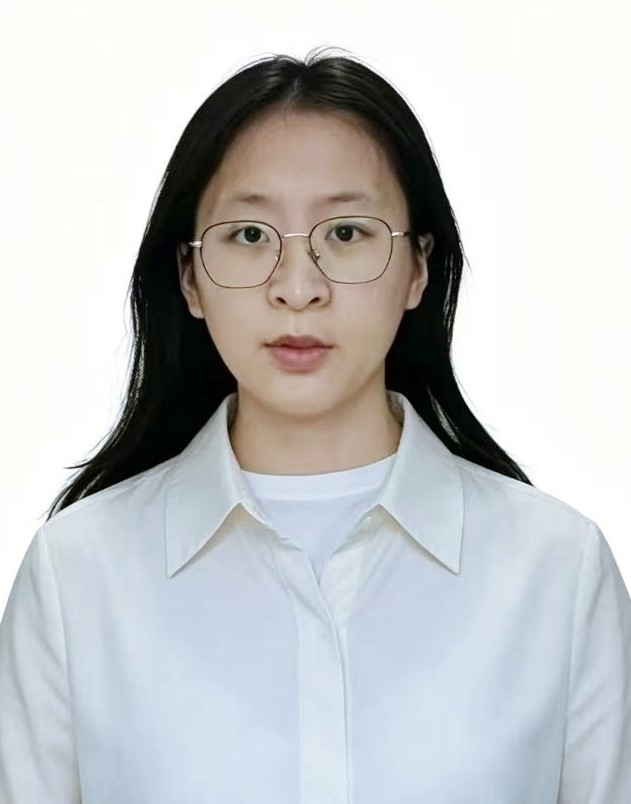}}]
{Shuyuan Guan}
is currently a Ph.D. student candidate jointly trained by the School of Cyberspace  Science, Harbin Institute of Technology and Zhongguancun Laboratory. Her research centers on the modeling and performance analysis of superconducting circuits. Before that, she received her bachelor’s and master’s degrees from Harbin Engineering University in 2022 and 2025, respectively.
\end{IEEEbiography}

\begin{IEEEbiography}[{\includegraphics[width=1in,height=1.25in,clip,keepaspectratio]{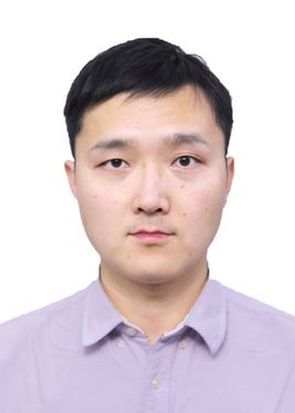}}]{Feng Wu}
   is an Associate Researcher at the Zhongguancun Laboratory. Prior to this, he served as a research scientist at the DAMO Quantum Laboratory from 2019 to 2024, following postdoctoral research at the University of California, Santa Cruz. He earned his PhD from Peking University in 2016. His current research focuses on the design, physical analysis, and verification of superconducting quantum computing processors.
\end{IEEEbiography}

\begin{IEEEbiography}
[{\includegraphics[width=1in,height=1.25in,clip,keepaspectratio]{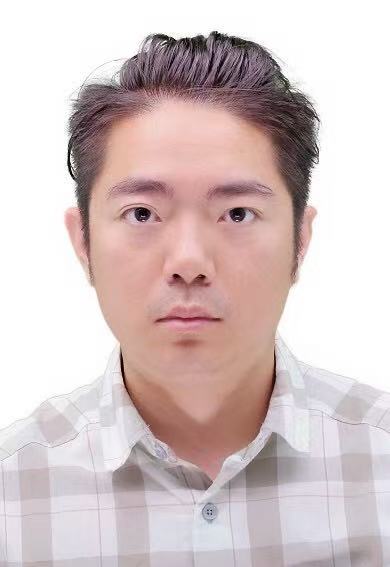}}]
{Xiaohang Zhang}
is the director of the Nanofabrication Facility at i-BRAIN and an adjunct professor at Zhejiang University, as well as a guest research scientist at the National Astronomical Observatories, Chinese Academy of Sciences, and ShanghaiTech University. He previously worked at the National Institute of Standards and Technology (NIST) in the United States, Alibaba Quantum Lab, and Zhejiang Laboratory. Dr. Zhang’s expertise spans nanofabrication, quantum sensors, superconducting devices, cryogenic systems, and advanced astronomical instrumentation. He is currently focused on developing transformative brain-computer interfaces (BCIs) that blur the distinction between electronics and the brain.
\end{IEEEbiography}

\begin{IEEEbiography}
[{\includegraphics[width=1in,height=1.25in,clip,keepaspectratio]{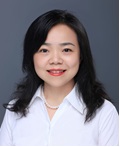}}]
{Qiong Li}
is a professor/doctoral supervisor, and the Director of the Information Countermeasures Technology Institute of the School of Cyberspace Science, Harbin Institute of Technology. Her research interests include information security, quantum cryptography, etc. She has completed more than 20 scientific research projects as the project leader; published more than 70 academic papers; authorized more than 40 invention patents.
\end{IEEEbiography}

\begin{IEEEbiography}
[{\includegraphics[width=1in,height=1.25in,clip,keepaspectratio]{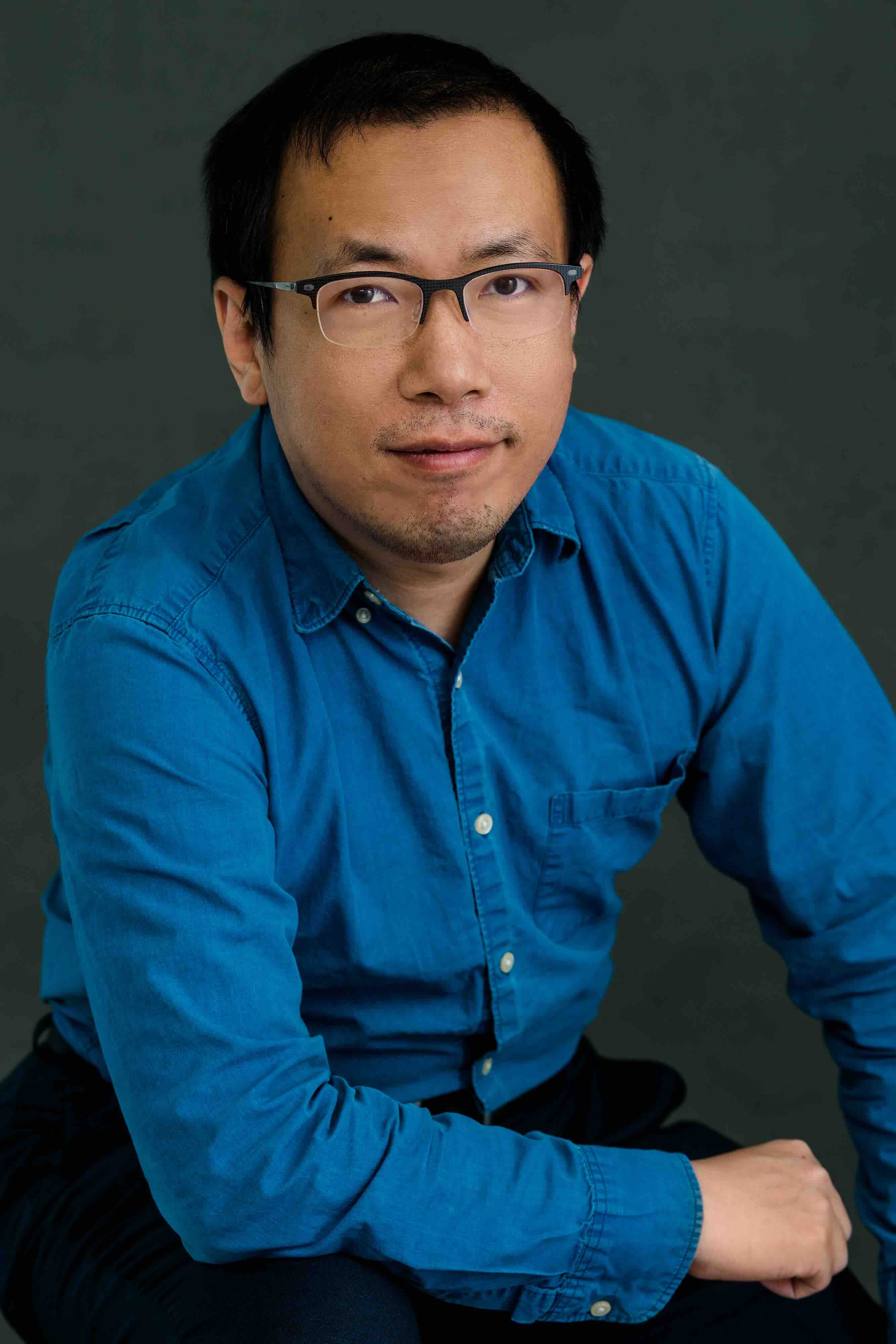}}]
{Jianxin Chen}
is a tenured associate professor in the Department of Computer Science and Technology at Tsinghua University. His recent research focuses on the system design and implementation of quantum computers, as well as quantum error correction and fault tolerance. He earned his Bachelor of Engineering (2005) and Doctor of Philosophy (2010) degrees in Computer Science from Tsinghua University. Subsequently, he conducted postdoctoral research at the Institute for Quantum Computing, University of Waterloo / University of Guelph (Canada), and the Joint Center for Quantum Information and Computer Science (QuICS), University of Maryland, College Park (USA). Prior to joining Tsinghua University in March 2025, he led R\&D efforts for both the systems team and the North American team at the Alibaba Quantum Laboratory.
\end{IEEEbiography}

\begin{IEEEbiography}
[{\includegraphics[width=1in,height=1.25in,clip,keepaspectratio]{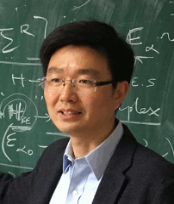}}]
{Xin Wan}
is a professor in Zhejiang Institute of Modern Physics at Zhejiang University. He received his Ph.D. degree in electrical engineering from Princeton University in 2000. He conducted postdoctoral research at the National High Magnetic Field Laboratory in Tallahassee and the Institute of Nanotechnology at now Karlsruhe Institute of Technology before joining Zhejiang University. His research focuses on condensed matter theory, in particular in two-dimensional electron systems.
\end{IEEEbiography}

\begin{IEEEbiography}
[{\includegraphics[width=1in,height=1.25in,clip,keepaspectratio]{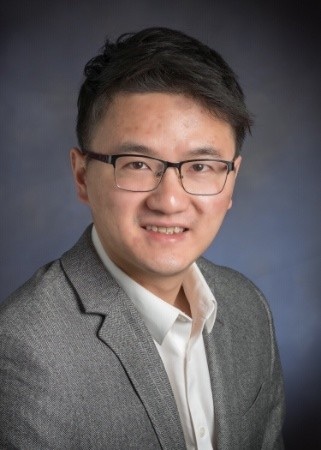}}]
{Tian Xia}
received the B.S. degree from Zhejiang University, Hangzhou, China and the University of New South Wales, Sydney, Australia in 2011, and Ph.D. degree in electrical engineering from the University of Illinois, Urbana-Champaign  on computational electromagnetics in 2018. After his PhD, he has been a hardware engineer at Apple Corporation, Cupertino, US. From 2020 to 2024, he worked as a quantum scientist at DAMO quantum laboratory, Hangzhou, China. He is the recipient of the Best Student Paper Award in Progress in PIERS 2016, and the recipient of the Y. T. Lo Outstanding Research Award in 2018. He is interested in the simulation and optimization algorithms for the physical design of the superconducting quantum chip. 
\end{IEEEbiography}

\begin{IEEEbiography}[{\includegraphics[width=1in,height=1.25in,clip,keepaspectratio]{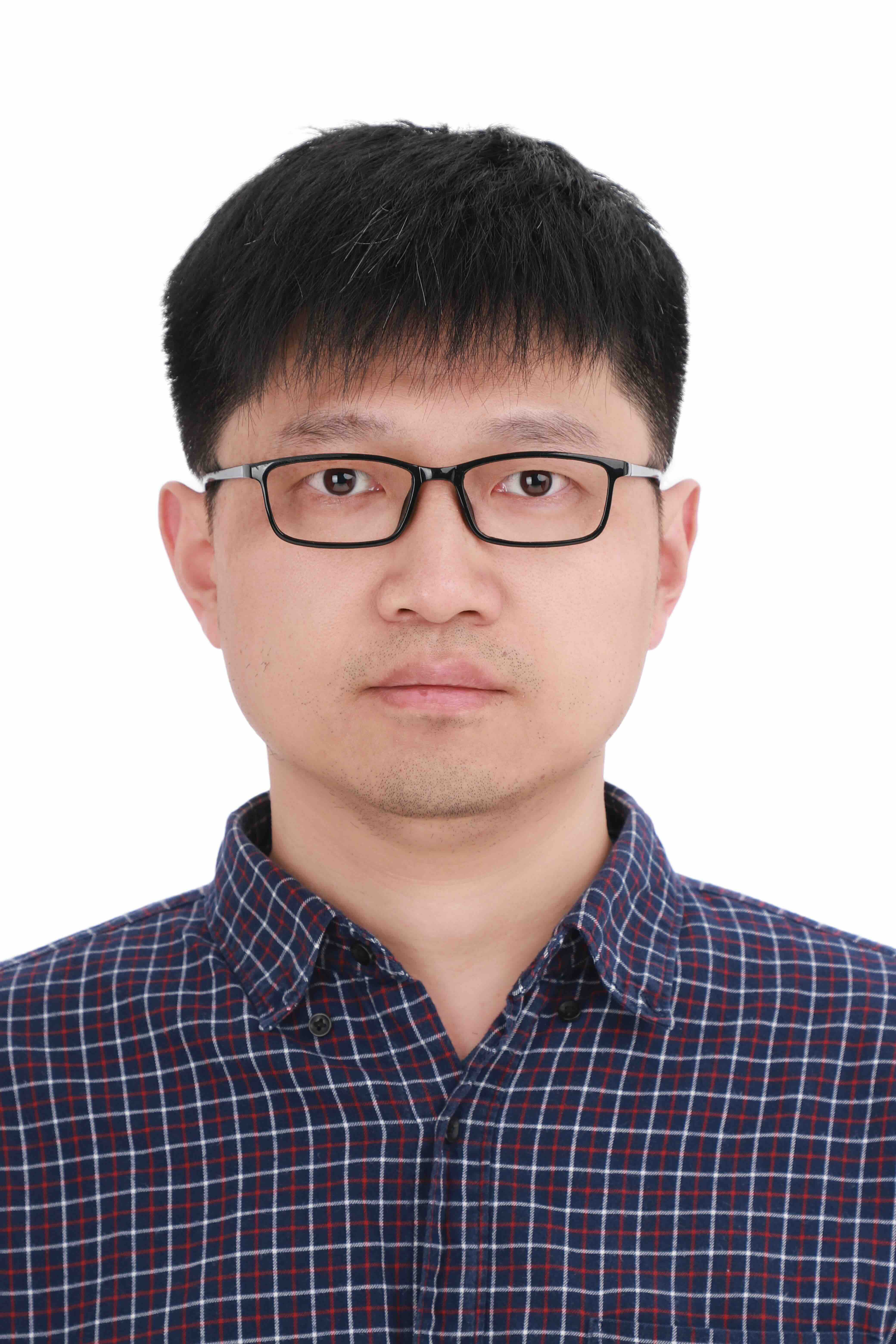}}]{Hui-Hai Zhao}
is a researcher at the Zhongguancun Laboratory. His recent research focuses on building design automation software toolchain and developing scalable, high-precision superconducting quantum hardware. He received his BSc in Physics from Yuanpei College of Peking University in 2007, and his PhD in condensed matter physics from Institute of Physics, Chinese Academy of Sciences in 2013. Subsequently, he conducted postdoctoral research at the University of Tokyo. In 2017, he joined the Institute of Physical and Chemical Research (RIKEN) in Japan as a research scientist. Prior to joining the Zhongguancun Laboratory in April 2024, he led R\&D efforts for the design team at the Alibaba Quantum Laboratory.
\end{IEEEbiography}

\vfill


\end{document}